\documentclass[aps,pre,reprint,showpacs,eqsecnum]{revtex4-1}
\usepackage{amssymb,amsmath}
\usepackage{graphicx}
\usepackage{color}
\usepackage{enumerate}
\usepackage[unicode=true,pdfusetitle,  bookmarks=true,bookmarksnumbered=false,bookmarksopen=false,  breaklinks=false,pdfborder={0 0 0},backref=false,colorlinks=true]  {hyperref}
\hypersetup{linkcolor=blue,citecolor=red}

\newcommand{\hnn}[1]{\widehat{\boldsymbol #1}}

\newcommand\beq{\begin{equation}}
\newcommand\eeq{\end{equation}}
\newcommand\beqa{\begin{eqnarray}}
\newcommand\eeqa{\end{eqnarray}}
\def\bal#1\eal{\begin{align}#1\end{align}}
\newcommand{\nn}{\nonumber\\}
\newcommand{\hcs}{\infty}
\newcommand{\bbA}{\boldsymbol{\mathcal{A}}}
\newcommand{\bbB}{\boldsymbol{\mathcal{B}}}
\newcommand{\bbC}{\mathcal{C}}
\newcommand{\bbE}{\mathcal{E}}
\newcommand{\bnabla}{\boldsymbol{\nabla}}
\newcommand{\thetah}{\theta_\hcs}

\def\bc{\mathbf{v}}
\def\bt{\widetilde\beta}
\def\at{\widetilde\alpha}
\def\bg{\mathbf{g}}
\def\bk{\hnn{\sigma}}
\def\a{{\alpha }}
\def\b{{\beta }}
\def\be{\begin{equation}}
\def\ee#1{\label{#1}\end{equation}}
\def\d{\sigma }
\def\bu{\overline{\mathbf{g}}}
\def\bx{\mathbf{r}}
\def\buu{\mathbf{u}}
\def\bw{\boldsymbol{\omega}}
\def\bJ{\mathbf{Q}}
\def\Tt{T_t}
\def\Tr{T_r}
\def\zt{\zeta_t}
\def\zr{\zeta_r}
\def\lb{\label}
\newcommand{\ben}{\begin{eqnarray}}
\newcommand{\een}{\end{eqnarray}}

\begin{document}
\title{Transport coefficients of a granular gas of inelastic rough hard spheres}
\author{Gilberto M.  Kremer}
\email{kremer@fisica.ufpr.br}
\affiliation{Departamento de F\'{\i}sica, Universidade Federal do Paran\'a, Curitiba, Brazil}

\author{Andr\'{e}s Santos}
\email{andres@unex.es}
\homepage{http://www.unex.es/eweb/fisteor/andres/}

\author{Vicente Garz\'{o}}
\email{vicenteg@unex.es}
\homepage{http://www.unex.es/eweb/fisteor/vicente/}
\affiliation{Departamento de F\'{\i}sica and Instituto de Computaci\'on Cient\'ifica Avanzada (ICCAEx), Universidad de Extremadura, E-06071 Badajoz, Spain}


\begin{abstract}
The Boltzmann equation for inelastic and rough hard spheres is considered as a model of a dilute granular gas. In this model, the collisions are characterized by constant coefficients of normal and tangential restitution and hence the translational and rotational degrees of freedom are coupled. A normal solution to the Boltzmann equation is obtained by means of the Chapman--Enskog method for states near the homogeneous cooling state. The analysis is carried out to first order in the spatial gradients of the number density, the flow velocity, and the granular temperature. The constitutive equations for the momentum and heat fluxes and for the cooling rate are derived, and the associated transport coefficients are expressed in terms of the solutions of linear integral equations. For practical purposes, a first Sonine approximation is used to obtain explicit expressions of the transport coefficients as nonlinear functions of both coefficients of restitution and the moment of inertia. Known results for purely smooth inelastic spheres and perfectly elastic and rough spheres are recovered in the appropriate limits.

\end{abstract}

\date{\today}
\maketitle

\section{Introduction}
\label{sec1}

As is well known, the prototypical model of a granular gas is a system composed of smooth, frictionless hard spheres which collide inelastically with a constant coefficient of normal restitution $0<\alpha\leq 1$ \cite{BP04,C90,G03}. In the dilute and moderately dense regimes, the microscopic description of the gas is given by the one-particle velocity distribution function $f$ obeying the (inelastic) Boltzmann and Enskog kinetic equations \cite{BP04,GS95,BDS97}. At a more phenomenological level, the gas can also be described by the Navier--Stokes--Fourier (NSF) hydrodynamic equations for the densities of mass, momentum, and energy with appropriate constitutive equations for the stress tensor, heat flux, and cooling rate. The Chapman--Enskog method \cite{CC70,FK72} bridges the gap between the kinetic and hydrodynamic descriptions, thus providing explicit expressions for the NSF transport coefficients in terms of the coefficient of normal restitution. This task was first accomplished in the quasismooth limit \cite{LSJC84,JR85b,SG98}, the results being subsequently extended to finite degree of inelasticity for monocomponent \cite{BDKS98,GD99} and multicomponent \cite{GD02,GDH07,GHD07,MGH12} granular gases.

In spite of the interest and success of the smooth hard-sphere model of granular gases, grains in nature are typically frictional, and hence energy transfer between the translational and rotational degrees of freedom occurs upon particle collisions. The simplest model accounting for particle roughness (and thus including the particle angular velocity as an additional mechanical variable) neglects the effect of sliding collisions and is characterized by a constant coefficient of tangential restitution $\beta$ \cite{JR85}. This parameter ranges from $-1$ (perfectly smooth spheres) to $1$ (perfectly rough spheres). A more sophisticated model incorporates the Coulomb friction coefficient as a third collision constant, so that collisions become sliding beyond a certain impact parameter \cite{W93,FLCA94,GNB05,GNB05b}. On the other hand, this three-parameter model significantly complicates the kinetic description, while the simpler two-parameter $(\alpha,\beta)$ model captures the essential features of granular flows when particle rotations are relevant. This explains the wide use of the latter model in the literature
\cite{JR85,LS87,C89,L91,LB94,L95,GS95,L96,HZ97,ZTPSH98,ML98,LHMZ98,HHZ00,AHZ01,MHN02,CLH02,JZ02,PZMZ02,GNB05,%
GNB05b,Z06,BPKZ07,GA08,KBPZ09,SKG10,N11,SKS11,BB12,MDHEH13,VSK14}.

Needless to say, an important challenge is the derivation of the NSF hydrodynamic equations of a granular gas of inelastic \emph{rough} hard spheres, with explicit expressions for the transport coefficients as functions of $\alpha$ and $\beta$. Previous attempts have been restricted to nearly elastic collisions ($\alpha\lesssim 1$) and  either nearly smooth particles ($\beta\gtrsim -1$) \cite{JR85,GNB05,GNB05b} or nearly perfectly rough particles ($\beta\lesssim 1$) \cite{JR85,L91}. The goal of this paper is to  uncover the whole range of values of the two coefficients of restitution $\alpha$ and $\beta$ and derive explicit expressions for the NSF transport coefficients of a \emph{dilute} granular gas beyond the above limiting situations.

In the case of conventional gases (i.e., when energy is conserved upon collisions), the set of hydrodynamic variables is  related to densities of conserved quantities, namely, the particle density $n$ (conservation of mass), the flow velocity $\mathbf{u}$ (conservation of momentum), and the temperature $T$ (conservation of energy). If the particles are perfectly elastic and rough ($\alpha=\beta=1$), what is conserved is the sum of the translational and the rotational kinetic energies and thus the granular temperature $T$ has  translational ($T_t$) and  rotational ($T_r$) contributions \cite{P22,MSD66,CC70}. Moreover, since the angular velocity of the particles is not a collisional invariant, the mean spin $\boldsymbol{\Omega}$ is not included either in the set of hydrodynamic variables or in the definition of the rotational contribution ($\Tr$) to the temperature \cite{P22,MSD66,CC70}.

For granular gases, although the total kinetic energy is dissipated by collisions,   the granular temperature is typically included as a hydrodynamic field  in most studies, and, consequently, a sink term appears in the corresponding balance equation. For nearly smooth spheres ($\beta\gtrsim -1$), some authors \cite{JR85,GNB05,GNB05b} have chosen (in addition to $n$ and $\mathbf{u}$) the two partial contributions $\Tt$ and $\Tr$ to the temperature, as well as the mean spin $\boldsymbol{\Omega}$,  as hydrodynamic variables. On the other hand, in this paper we will choose as hydrodynamic fields for dissipative gases the same as in conservative systems, i.e., $n$, $\mathbf{u}$, and $T$. In this way, the hydrodynamic description encompasses the conservative gases as special limits. The advantage of the choice of the set $\{n,\mathbf{u},T\}$ instead of the set $\{n,\mathbf{u},\boldsymbol{\Omega},\Tt,\Tr\}$ is analogous to the advantage of the set $\{n_1,n_2,\mathbf{u},T\}$ instead of $\{n_1,n_2,\mathbf{u}_1,\mathbf{u}_2,T_1,T_2\}$ in a binary mixture of inelastic smooth hard spheres, as discussed in Refs.\ \cite{GMD06,DB11}.

The plan of the paper is as follows. Section \ref{sec2} is devoted to the definition of the model of inelastic rough hard spheres and their description in the low-density regime by means of the Boltzmann equation. The  exact balance equations for the densities of mass, momentum, and energy are obtained from the Boltzmann equation, and the associated fluxes of momentum and energy, as well as the cooling rate, are identified.
The so-called homogeneous cooling state (HCS) is studied in Sec.\ \ref{sec3}. Special attention is paid to the time evolution of the mean spin $\boldsymbol{\Omega}$.  While the temperature ratio $\Tr/\Tt$ reaches a well-defined value for long times, the ratio $I\Omega^2/\Tr$ ($I$ being the moment of inertia) decays to zero with a characteristic time typically smaller {(except near $\beta= -1$)} than the relaxation time of the temperature ratio (see Fig.\ \ref{fig1}). This clearly justifies the exclusion of $\boldsymbol{\Omega}$ as a hydrodynamic field. Next, the Chapman--Enskog method is applied in Sec.\ \ref{sec4} to derive the linear integral equations for the velocity-dependent functions characterizing the distribution function to first-order in the hydrodynamic gradients. In Sec.\ \ref{sec5},  the NSF transport coefficients (shear and bulk viscosities, thermal conductivity, Dufour-like coefficient, and cooling rate transport coefficient) are expressed in terms of integrals involving the solutions of the linear integral equations. Next, a first Sonine polynomial approximation is used to obtain practical results from this formulation, thus providing explicit forms for the transport coefficients as nonlinear functions of both $\alpha$ and $\beta$ (see Table \ref{table1}). Some technical details of the calculations are relegated to the Appendix.
The results are discussed in Sec.\ \ref{sec6}, where known expressions in the limiting cases of inelastic smooth spheres ($\alpha<1$, $\beta=-1$) and perfectly elastic and rough spheres ($\alpha=\beta=1$) are recovered. The intricate dependence of the transport coefficients on both coefficients of restitution is illustrated by some representative cases. Finally, the paper closes with some concluding remarks in Sec.\ \ref{sec7}.

{\renewcommand{\arraystretch}{2}
\begin{table}
   \caption{Summary of explicit expressions.}\label{table1}
\begin{ruledtabular}
\begin{tabular}{l}
$\displaystyle{\widetilde\alpha={1+\alpha\over2}}$, $\displaystyle{\bt={1+\beta\over2}{\kappa\over\kappa+1}}$\\
$\displaystyle{\frac{\Tt^{(0)}}{T}=\tau_t={2\over1+\thetah}}$, $\displaystyle{\frac{\Tr^{(0)}}{T}=\tau_r={2\thetah\over1+\thetah}}$\\
$\displaystyle{\thetah=\sqrt{1+h^2}+h}$, $\displaystyle{h\equiv \frac{(1+\kappa)^2}{2\kappa(1+\beta)^2}\left[{1-\alpha^2}-(1-\beta^2)\frac{1-\kappa}{1+\kappa}\right]}$\\
$\displaystyle{\nu=\frac{16}{5}\sigma^2 n\sqrt{\pi \tau_tT/m}}$\\
$\displaystyle{\frac{\zeta^{(0)}}{\nu}=\zeta^*=\frac{5}{12}\frac{1}{1+\thetah}\left[1-\alpha^2+({1-\beta^2})
\frac{\thetah+\kappa}{1+\kappa}\right]}$\\[2mm]
\hline
 $\displaystyle{\eta=\frac{n\tau_t T}{\nu}\frac{1}{\nu_\eta^*-\frac{1}{2}\zeta^*}}$\\
$\displaystyle{\eta_b=\frac{n\tau_t\tau_r T}{\nu}\gamma_E}$\\
$\displaystyle{\lambda={\tau_t\lambda_t+\tau_r\lambda_r}}$, $\displaystyle{\lambda_t=\frac{5}{2}\frac{n\tau_t T}{m\nu}\gamma_{A_t}}$, $\displaystyle{\lambda_r=\frac{3}{2}\frac{n\tau_t T}{m\nu}\gamma_{A_r}}$\\
 $\displaystyle{\mu={\mu_t+\mu_r}}$, $\displaystyle{\mu_t=\frac{5}{2}\frac{\tau_t^2 T^2}{m\nu}\gamma_{B_t}}$, $\displaystyle{\mu_r=\frac{3}{2}\frac{\tau_t \tau_r T^2}{m\nu}\gamma_{B_r}}$\\
$\displaystyle{\xi={\frac{1}{2}\left(\tau_t\xi_t+\tau_r\xi_r\right)}=\gamma_E\Xi}$, $\displaystyle{\xi_t=\gamma_E\Xi_t}$, $\displaystyle{\xi_r=\gamma_E\Xi_r}$\\[1mm]
\hline
$\displaystyle{\nu_\eta^*=(\at+\bt)(2-\at-\bt)+\frac{\bt^2\thetah}{6\kappa}}$\\
$\displaystyle{\gamma_E=\frac{2}{3}({\Xi_t-\Xi_r-\zeta^*})^{-1}}$\\
$\displaystyle{\Xi_t=\frac{5}{8}\tau_r\Big[1-\a^2+(1-\b^2)\frac{\kappa}{1+\kappa}-\frac{\kappa}{3}({\thetah-5})\left(\frac{1+\b}{1+\kappa}\right)^2
\Big]}$\\
$\displaystyle{\Xi_r=\frac{5}{8}\tau_t\frac{1+\beta}{1+\kappa}\left[\frac{\thetah-2}{3}(1-\beta)
+\frac{\kappa}{3}({\thetah-5})\frac{1+\b}{1+\kappa}\right]}$\\
$\displaystyle{\Xi=\frac{5}{16}\tau_t\tau_r\left[1-\a^2+(1-\b^2)\left(1+\frac{1}{3}\frac{\thetah-5}{1+\kappa}\right)\right]}$\\
$\displaystyle{\gamma_{A_t}=\frac{Z_r-Z_t-2\zeta^*}{\left(Y_t-2\zeta^*\right)\left(Z_r-2\zeta^*\right)-Y_r Z_t}}$\\
$\displaystyle{\gamma_{A_r}=\frac{Y_t-Y_r-2\zeta^*}{\left(Y_t-2\zeta^*\right)\left(Z_r-2\zeta^*\right)-Y_r Z_t}}$\\
$\displaystyle{\gamma_{B_t}=\zeta^*\frac{\gamma_{A_t}\left(Z_r-\frac{3}{2}\zeta^*\right)-\gamma_{A_r}Z_t}
{\left(Y_t-\frac{3}{2}\zeta^*\right)\left(Z_r-\frac{3}{2}\zeta^*\right)-Y_r Z_t}}$\\
$\displaystyle{\gamma_{B_r}=\zeta^*\frac{\gamma_{A_r}\left(Y_t-\frac{3}{2}\zeta^*\right)-\gamma_{A_t}Y_r}
{\left(Y_t-\frac{3}{2}\zeta^*\right)\left(Z_r-\frac{3}{2}\zeta^*\right)-Y_r Z_t}}$\\
$\displaystyle{Y_t=\frac{41}{12}\left(\at+\bt\right)-\frac{33}{12}\left(\at^2+\bt^2\right)-\frac{4}{3}\at\bt-\frac{7\thetah}{12}
\frac{\bt^2}{\kappa}}$\\
$\displaystyle{Z_t=-\frac{5\thetah}{6}
\frac{\bt^2}{\kappa}}$\\
$\displaystyle{Y_r=\frac{25}{36}\frac{\bt}{\kappa}\left(1-3\frac{\bt}{\thetah}-\frac{\bt}{\kappa}\right)}$\\
$\displaystyle{Z_r=\frac{5}{6}\left(\at+\bt\right)+\frac{5}{18}\frac{\bt}{\kappa}\left(7-3\frac{\bt}{\kappa}-6{\bt}-
 4{\at}\right)}$\\
     \end{tabular}
 \end{ruledtabular}
 \end{table}
}

\section{Granular gas of inelastic rough hard spheres: Boltzmann description}
\label{sec2}

\subsection{Collision rules}

Let  $(\bc, \bc_1)$ and $(\bw, \bw_1)$ denote the linear and the angular  precollisional velocities, respectively, of two rough spherical particles with the same mass $m$, diameter $\d$, and moment of inertia $I$, while $(\bc^\prime, \bc_1^\prime)$ and $(\bw^\prime,\bw_1^\prime)$ correspond to their postcollisional velocities. The pre- and postcollisional velocities are related by
\begin{subequations}
\label{1,2}
\bal
\label{1}
 m \bc^\prime=m\bc -\bJ,\quad &I \bw^\prime=I\bw -{\d\over2}\,\bk\times\bJ,\\
 \label{2}
 m \bc_1^\prime=m\bc_1 +\bJ,\quad
 &I \bw_1^\prime=I\bw_1 -{\d\over2}\,\bk\times\bJ,
 \eal
\end{subequations}
where ${\bJ}$ denotes the impulse exerted by the unlabeled particle on the labeled one and $\bk$ is the unit collision vector  joining the centers of the two colliding spheres and pointing from the center of the {unlabeled} particle to the center of the {labeled one}. Furthermore, the relationship between the center-of-mass relative velocities $(\bg=\bc-\bc_1,\bg'=\bc'-\bc_1')$ and the  relative velocities $(\bu,\bu')$ of the points of the spheres which are in contact during a binary encounter are
\begin{subequations}
\label{3}
\beq
 \bu=\bg-{\d\over2}\bk\times(\bw+\bw_1),
 \label{3a}
 \eeq
 \beq
 \label{3b}
 \bu^\prime=\bg^\prime-{\d\over2}\bk\times(\bw^\prime+\bw_1^\prime).
 \eeq
\end{subequations}

Combining Eqs.\ \eqref{1,2} and \eqref{3}, one obtains
\bal
\bu'=&\bu-\frac{2}{m}\bJ+\frac{2}{m\kappa}\bk\times(\bk\times\bJ)\nn
=&\bu-\frac{2}{m}{\kappa+1\over \kappa}\bJ+\frac{2}{m\kappa}(\bk\cdot\bJ)\bk,
\label{3bg}
\eal
where in the second step use has been made of the mathematical property $\bk\times(\bk\times\bJ)=(\bk\cdot\bJ)\bk-\bJ$ and
\beq
\kappa={4I\over m\d^2}
\eeq
is a dimensionless moment of inertia, which may vary from zero to a maximum value of $2/3$, the former corresponding to a concentration of the mass at the center of the sphere, while the latter corresponds to a concentration of the mass on the surface of the sphere. The value $\kappa=2/5$ refers to a uniform distribution of the mass in the sphere.

The inelastic collisions of rough spherical particles are characterized by the relationships
 \be
\bk\cdot\bu^\prime=-\alpha(\bk\cdot\bu),\quad
\bk\times\bu^\prime=-\beta(\bk\times\bu),
 \ee{4}
where $0<\alpha\leq1$ and $-1\leq\beta\leq1$ are the normal and tangential restitution coefficients, respectively. For an elastic collision of perfectly smooth spheres one has $\alpha=1$ and $\beta=-1$, while $\alpha=1$ and $\beta=1$ for an elastic encounter of perfectly rough spherical particles.

{}Insertion of Eq.\ \eqref{3bg} into Eq.\ \eqref{4} gives $\bk\cdot\bJ=m\at(\bk\cdot\bu)$ and $\bk\times\bJ=m\bt(\bk\times\bu)$, where the following abbreviations have been introduced:
 \beq\label{7}
 \widetilde\alpha\equiv{1+\alpha\over2},\quad \widetilde\beta\equiv{1+\beta\over2}{\kappa\over\kappa+1}.
 \eeq
Therefore, the impulse can be expressed as
 \bal
\bJ=&{m\widetilde\alpha}(\bk\cdot\bu)\bk-m\widetilde\beta\bk\times(\bk\times\bu)\nn
=&{m\widetilde\alpha}(\bk\cdot\bg)\bk-m\widetilde\beta\bk\times\left(\bk\times\bg+\d{\bw+\bw_1\over 2}\right).
\label{6}
 \eal

Equations \eqref{1,2} and (\ref{6})  express the  postcollisional velocities  in terms of the precollisional velocities and of the collision vector \cite{note_14_05_1}. {}From these results it is easy to obtain that the change of the total (translational plus rotational) kinetic energy reads
 \bal
 \Delta K=&{m\over2}\left(v^{\prime2}+v_1^{\prime2}-v^2-v_1^2\right)+{I\over2}\left(\omega^{\prime2}+\omega_1^{\prime2}-\omega^2-\omega_1^2\right)\nn
   =&-m{1-\beta^2\over4}{\kappa\over\kappa+1}\left[\bk\times\left(\bk\times\bg+\d{\bw+\bw_1\over 2}\right)\right]^2\nn
   &-m{1-\alpha^2\over4}(\bk\cdot\bg)^2.
  \label{Delta_E}
 \eal
The right-hand side  vanishes for elastic collisions of perfectly smooth spheres $(\alpha=1,\beta=-1)$ and for  elastic collisions of perfectly rough spherical particles  $(\alpha=1, \beta=1)$. In those cases the total energy is conserved in a collision.

Apart from the linear momentum, the angular momentum is conserved, namely,
\bal
m (\bx\times \bc +\bx_1\times \bc_1)+I(\bw+\bw_1)=&m (\bx\times \bc' +\bx_1\times \bc_1')\nn
&+I(\bw'+\bw_1'),
\label{bal_ang}
\eal
where $\bx$ and $\bx_1=\bx+\sigma\bk$ are the position vectors of the two colliding particles.

\subsection{Boltzmann equation}
A direct encounter is characterized by the precollisional velocities $(\bc, \bc_1; \bw,
 \bw_1)$, by the postcollisional velocities $(\bc^\prime, \bc_1^\prime; \bw^\prime,
 \bw_1^\prime)$, and by the collision vector $\bk$. For a restitution encounter
  the pre- and postcollisional velocities are denoted by $(\bc^\ast, \bc_1^\ast; \bw^\ast,
 \bw_1^\ast)$ and $(\bc, \bc_1; \bw, \bw_1)$, respectively,  and the collision vector by $\bk^\ast=-\bk$.
It is easy to verify  the relationship $\bk^\ast\cdot\bg=-\alpha(\bk^\ast\cdot\bg^\ast)=-\bk\cdot\bg$.
The modulus of the Jacobian of the  transformation $(\bc^\ast, \bc_1^\ast; \bw^\ast,
 \bw_1^\ast)\to(\bc, \bc_1; \bw, \bw_1)$ is given by
\beq
\left|{\partial(\bc^\ast, \bc_1^\ast; \bw^\ast,
 \bw_1^\ast)\over\partial(\bc, \bc_1; \bw, \bw_1)}\right|={1\over\alpha\beta^2}
 \label{13}
 \eeq
Thus,
 \be
 (\bk^\ast\cdot\bg^\ast) d\bc^\ast d\bw^\ast d\bc_1^\ast d\bw_1^\ast={1\over\alpha^2\beta^2}
 (\bk\cdot\bg)d\bc d\bw d\bc_1 d\bw_1.
 \ee{14}

From Eq.\ (\ref{14}) we may infer that the Boltzmann equation for granular gases of rough spherical particles without external forces and torques is given by
\begin{subequations}
  \label{15}
  \beq
{\partial f \over \partial t}+\bc\cdot\bnabla f=J[f,f],
\label{15a}
\eeq
 \beq
J[f,f]=\d^2\int d\bc_1\int d\bw_1\int_{+} d\bk\,(\bk\cdot\bg)\left({f_1^\ast f^\ast\over\alpha^2\beta^2} -f_1f\right),
\label{15b}
 \eeq
\end{subequations}
where  $f(\bx,\bc, \bw, t)$ is the one-particle distribution function in the phase space spanned by the positions and the linear  and angular
velocities of the particles.
As usual, in Eq.\ \eqref{15b} the notation $f_1=f(\bc_1, \bw_1)$, $f^*=f(\bc^*, \bw^*)$, $f_1^*=f(\bc_1^*, \bw_1^*)$ has been employed. Also, the subscript ($+$) in the integration over $\bk$ denotes the constraint $\bk\cdot\bg>0$.

The so-called transfer equation  is obtained from the multiplication of the Boltzmann equation by an arbitrary function $\psi(\bx,\bc,\bw,t)$ and integration of the resulting equation over all values of the velocities $\bc$ and $\bw$, yielding
\beq
\label{16}
\partial_t \left(n\langle\psi\rangle\right) +\bnabla\cdot \left(n \langle \bc\psi\rangle\right)-n\langle (\partial_t+\bc\cdot\bnabla)\psi\rangle=
\mathcal{J}[\psi|f,f],
\eeq
where
\beq
n(\bx,t)=\int d\bc\int d\bw\, f(\bx,\bc,\bw,t)
\label{n}
\eeq
is the local number density,
\beq
\langle\psi\rangle ={1\over n(\bx,t)}\int d\bc\int d\bw\, \psi(\bx,\bc,\bw,t)f(\bx,\bc,\bw,t)
\label{avpsi}
\eeq
is the local average value of $\psi$, and
\bal
\label{16J}
\mathcal{J}[\psi|f,f]\equiv&\int d\bc\int d\bw\,\psi(\bx,\bc,\bw,t)J[f,f]\nn
=& {\d^2\over2}\int d\bc\int d\bw\int d\bc_1\int d\bw_1\int_+ d\bk\,(\bk\cdot\bg)\nn
&\times\left(\psi_1'+\psi'-\psi_1-\psi\right)f_1f
\eal
is the collisional production term of $\psi$.
In the second step  of Eq.\ \eqref{16J} we have used the relationship (\ref{14}) and the standard symmetry properties of the collision term.

 \subsection{Balance equations}

A macroscopic description of a rarefied granular gas of rough spheres can be characterized by the following basic fields: the particle number density
$n({\bx},t)$ [see Eq.\ \eqref{n}], the hydrodynamic flow velocity
$\buu({\bx},t)$, and the granular temperature $T(\bx,t)$. The two latter quantities are defined as
\beq
\buu=\langle \bc\rangle, \quad T={1\over2}(\Tt+\Tr),
\label{u,T}
\eeq
where
\beq
\Tt={m\over3}\langle V^2\rangle,\quad \Tr={I\over3}\langle \omega^2\rangle
\label{Tt,Tr}
\eeq
are the (partial) translational and rotational temperatures, respectively, and the averages $\langle \cdots\rangle$ are defined by Eq.\ \eqref{avpsi}. In Eq.\ \eqref{Tt,Tr}, $\mathbf{V}=\bc-\buu$ is the (translational) peculiar velocity. On the other hand, in the definition of $\Tr$ we have chosen not to refer the
angular velocities to the mean value
\beq
\boldsymbol{\Omega}=\langle \bw\rangle
 \label{Omega}
 \eeq
because the latter is not a conserved quantity. Had we defined the granular temperature as $\overline{T}=(\Tt+\overline{T}_r)/2$ with $\overline{T}_r={I\over3}\langle(\bw- \boldsymbol{\Omega})^2\rangle$, then $\overline{T}$ would not be a conserved quantity in the case of completely rough and elastic collisions ($\alpha=\beta=1$), even though $\Delta K=0$ in that case [see Eq.\ \eqref{Delta_E}]

The balance equations for the basic fields are obtained from the transfer equation (\ref{16})  with the following choices for the arbitrary function $\psi(\bx,\bc,\bw,t)$:
\begin{enumerate}
\item
 Balance of particle number density ($\psi=1$),
\beq
\label{18}
{\cal D}_tn+n\bnabla\cdot \buu=0.
\eeq
\item
Balance of momentum density ($\psi=m\bc$),
\beq
\label{19}
\rho{\cal D}_t \buu
+\bnabla\cdot \mathsf{P}
=0.
\eeq
\item
Balance of   temperature ($\psi=mV^2/2+I\omega^2/2$),
\beq
\label{20}
{\cal D}_t T +{1\over3n}\left(
\bnabla\cdot \mathbf{q}+\mathsf{P}:\bnabla \buu\right)+ T  \zeta=0.
\eeq
\end{enumerate}
In the above balance equations, $\rho=mn$ is the mass density, $\mathcal{D}_t=\partial_t+\buu\cdot\bnabla$ denotes the material time derivative, and the following quantities have been introduced: the pressure tensor
\beq
P_{ij}=\rho \langle V_iV_j\rangle,
\eeq
the heat flux vector
\beq
\mathbf{q}=\mathbf{q}_t+\mathbf{q}_r
\eeq
with
\beq
\mathbf{q}_t={\rho\over2}\langle V^2\mathbf{V}\rangle,\quad
\mathbf{q}_r={In\over2}\langle \omega^2\mathbf{V}\rangle,
\eeq
and the cooling rate
\beq
\label{zeta}
\zeta={\Tt\over 2T}\zt+{\Tr\over 2T}\zr
\eeq
with
\begin{subequations}
\label{ztrrot}
\beq
\label{ztr}
\zt=-\frac{m}{3n\Tt}\mathcal{J}[v^2|f,f],
\eeq
\beq
\label{zrot}
\zr=-\frac{I}{3n\Tr}\mathcal{J}[\omega^2|f,f].
\eeq
\end{subequations}

For further use, we note that the hydrostatic pressure $p$ is defined as one third of the trace of the pressure tensor, so that
\beq
p=n\Tt.
\eeq
Also, the balance equations for the partial temperatures are
\beq
\label{20t}
{\cal D}_t \Tt +{2\over3n}\left(
\bnabla\cdot \mathbf{q}_t+\mathsf{P}:\bnabla \buu\right)+ \Tt  \zt=0,
\eeq
\beq
\label{20r}
{\cal D}_t \Tr +{2\over3n}
\bnabla\cdot \mathbf{q}_r+ \Tr  \zr=0.
\eeq
Combination of Eqs.\ \eqref{20t} and \eqref{20r} yields Eq.\ \eqref{20}.

It is worth noting that the conservation of angular momentum, Eq.\ \eqref{bal_ang}, does not generate a balance equation with $\mathcal{J}[\psi|f,f]=0$ for the quantity $\psi= m \bx\times\bc+I\bw$ because of the difference $\bx_1-\bx=\sigma\bk$ between the centers of the two colliding spheres.

\section{Homogeneous cooling state}
\label{sec3}
Before considering the transport properties in inhomogeneous states, it is convenient to characterize the main properties of homogeneous states. In those cases, Eqs.\ \eqref{18} and \eqref{19} imply $n=\text{const}$ and $\buu=\text{const}$, while Eqs.\ \eqref{20t} and \eqref{20r} become
\beq
\label{20tt}
\partial_t\Tt + \Tt  \zt=0,\quad
\partial_t \Tr + \Tr  \zr=0.
\eeq
The exact forms for the cooling rates $\zt$ and $\zr$ cannot be determined, unless the distribution function $f(\bc,\bw,t)$ is known. Good estimates of those quantities can be obtained by assuming the approximation
\beq
f(\bc,\bw)\to  \left({m \over 2\pi \Tt}\right)^{3/2} e^{-mV^2/2\Tt}f_r(\bw),
\label{3.1}
\eeq
where
\beq
f_r(\bw)=\int d\bc\, f(\bc,\bw)
\eeq
is the marginal distribution of angular velocities.
Equation \eqref{3.1} can be justified by maximum-entropy arguments, except that the explicit expression of $f_r(\bw)$ does not need to be specified. Substitution of \eqref{3.1} into Eqs.\ \eqref{ztrrot} and evaluation of the collision integrals given by Eq.\ \eqref{16J} yields \cite{SKG10}
\bal
\zt=&\frac{5}{12}\Bigg[1-\alpha^2+\frac{\kappa}{1+\kappa}\left(1-\beta^2\right)-\frac{\kappa}{(1+\kappa)^2}\left(1+\beta\right)^2
\nn
&\times \theta\left(1+ X-{1\over\theta}\right)\Bigg]\nu,
\label{2.12}
\eal
\bal
\zr=&\frac{5}{12}\frac{1+\beta}{1+\kappa}\Bigg[(1-\beta){(1+X)}+\frac{\kappa}{1+\kappa}\left(1+\beta\right)
\nn
&\times\left(1+X-{1\over\theta}\right)\Bigg]\nu,
\label{2.13}
\eal
where
\beq
\theta\equiv {\Tr\over \Tt},\quad X\equiv {I\Omega^2\over 3\Tr},
\eeq
and
\beq
\nu\equiv\frac{16}{5}\sigma^2 n\sqrt{\pi \Tt/m}
\label{nu}
\eeq
is an effective collision frequency. Note that $X=1-\overline{T}_r/\Tr\leq 1$.
The total cooling rate is, according to Eq.\ \eqref{zeta},
\beq
\zeta=\frac{5}{12}\frac{1}{1+\theta}\left[1-\alpha^2+\frac{1-\beta^2}{1+\kappa}\theta\left(\frac{\kappa}{\theta}+1+X\right)\right]\nu.
\label{2.14}
\eeq

Since Eqs.\ \eqref{2.12} and \eqref{2.13} involve the norm $\Omega^2$ of the mean angular velocity, we need to complement Eq.\ \eqref{20tt} with the evolution equation for $\Omega^2$. By using again the approximation \eqref{3.1} one obtains \cite{SKG10}
\beq
\partial_t \Omega^2+2\zeta_\Omega \Omega^2=0,\quad \zeta_\Omega=\frac{5}{6}{1+\beta\over 1+\kappa}\nu.
\label{zetaO}
\eeq
By introducing the time variable $s(t)=\int_0^t dt'\,\nu(t')$, which measures the (nominal) number of collisions per particle from the initial time to time $t$, the solution to Eq.\ \eqref{zetaO} is
\beq
\Omega^2(s)=\Omega^2(0)e^{-2\zeta_\Omega^* s},\quad \zeta_\Omega^*\equiv \zeta_\Omega/\nu.
\label{Omega2}
\eeq
This allows us to solve the set of coupled equations \eqref{20tt} to obtain the time dependence of the temperatures $\Tt(s)$ and $\Tr(s)$. Both quantities asymptotically decrease in time due to energy dissipation. On the other hand, the relevant quantity is the temperature ratio $\theta(s)$. Analogously, rather than the time decay of $\Omega^2(s)$, and since $\Tr(s)$ also tends to decay in time, the relevant quantity is the ratio $X(s)$. The evolution equations for both quantities are
\begin{subequations}
\beq
\partial_s \theta+(\zeta_r^*-\zeta_t^*)\theta=0,
\eeq
\beq
 \partial_s X+(2\zeta_\Omega^*-\zeta_r^*)X=0,
\eeq
\end{subequations}
where $\zeta_t^*\equiv\zt/\nu$ and $\zeta_r^*\equiv\zr/\nu$. Since
\bal
2\zeta_\Omega^*-\zeta_r^*=&\frac{5}{12}\frac{1+\beta}{(1+\kappa)^2}\Big[3-X+\beta(1+X)+2\kappa(1-X)\nn
&+\kappa\frac{1+\beta}{\theta}\Big]
\eal
is positive definite, it follows that $X\to 0$ monotonically, no matter the initial condition. On the other hand, the  evolution equation for $\theta$ admits a nonzero stationary solution given by the condition $\zeta_r^*=\zeta_t^*$. Such a solution is
\beq
\theta_\hcs=\sqrt{1+h^2}+h
\label{1.8}
\eeq
with
\beq
h\equiv \frac{1+\kappa}{2\kappa(1+\beta)}\left[(1+\kappa)\frac{1-\alpha^2}{1+\beta}-(1-\kappa)(1-\beta)\right].
\label{1.9}
\eeq

\begin{figure}
\includegraphics[width=8cm]{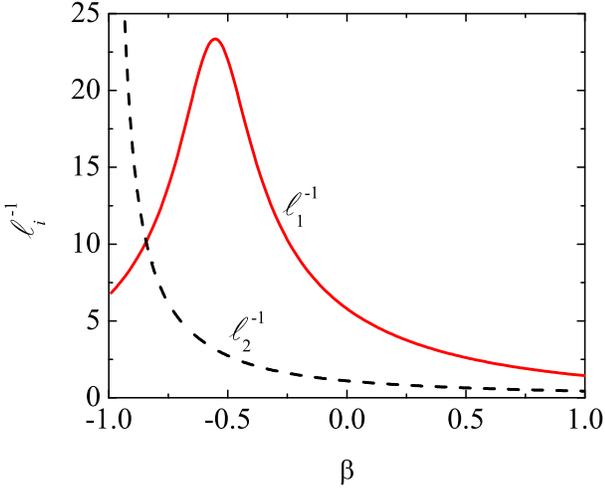}
  \caption{{(Color online) Plot of the characteristic relaxation times $\ell_1^{-1}$ and $\ell_2^{-1}$ as functions of $\beta$ for $\kappa=\frac{2}{5}$ and $\alpha=0.8$}.}
  \label{nfig1}
\end{figure}

A standard linear stability analysis of the stationary solution $X=0$ and $\theta=\theta_\hcs$ shows that the two associated eigenvalues are
\bal
\ell_{1}=&\theta_\hcs\left.\frac{\partial (\zeta_r^*-\zeta_t^*)}{\partial\theta}\right|_{\theta=\theta_\hcs,X=0}\nn
=&\frac{5}{12}\kappa\left(\frac{1+\beta}{1+\kappa}\right)^2\frac{1+\theta_\hcs^2}{\theta_\hcs},
\eal
\bal
\ell_{2}=&\left.{2\zeta_\Omega^*-\zeta_r^*}\right|_{\theta=\theta_\hcs,X=0}\nn
=&\frac{5}{12}\frac{1+\beta}{(1+\kappa)^2}\left(3+\beta+2\kappa+\kappa\frac{1+\beta}{\theta_\hcs}\right).
\eal
As expected, both eigenvalues are positive definite, what proves the stability of the stationary solution $(\theta,X)=(\theta_\hcs,0)$  against homogeneous perturbations.
{The time evolution of $X(s)$ is governed by the eigenvalue $\ell_2$ only, so that the relaxation time (in units of number of collisions per particle) of $X(s)$ is $s\sim \ell_2^{-1}$. As for $\theta(s)$, its relaxation time is  $s\sim \ell_1^{-1}$ if $X(0)=0$ and $s\sim \max(\ell_1^{-1},\ell_2^{-1})$ if $X(0)\neq 0$. Figure \ref{nfig1} shows the dependence of both relaxation times on roughness for the representative case $\alpha=0.8$. Except in a  narrow roughness region adjacent to $\beta=-1$, one has $\ell_1^{-1}>\ell_2^{-1}$, so that $X(s)$ relaxes more rapidly than $\theta(s)$. }

\begin{figure}
\includegraphics[width=8cm]{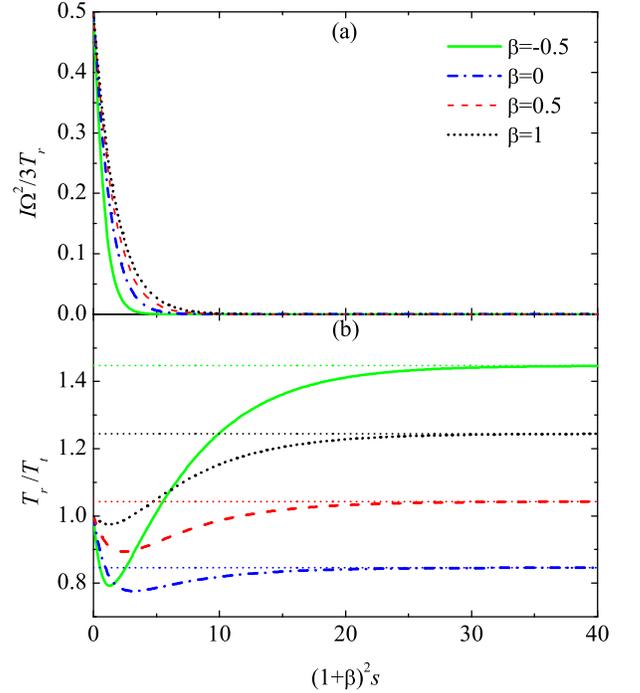}
  \caption{(Color online) Plot of (a) $X=I\Omega^2/3\Tr$ and (b) $\theta=\Tr/\Tt$ as functions of the number of collisions per particle ($s$) scaled by $(1+\beta)^{-2}$ for $\kappa=\frac{2}{5}$, $\alpha=0.8$, and $\beta=-0.5$, $0$, $0.5$, and $1$, starting from the initial condition $\theta(0)=1$, $X(0)=\frac{1}{2}$. The horizontal lines in panel (b) correspond to the respective stationary values $\theta_\hcs$.}
  \label{fig1}
\end{figure}

As an illustration of the evolution of $\theta(s)$ and $X(s)$, Fig.\ \ref{fig1} shows both quantities for $\alpha=0.8$ and $\beta=-0.5$, $0$, $0.5$, and $1$, starting from the initial condition $\theta(0)=1$, $X(0)=\frac{1}{2}$. It is quite apparent that $\theta$ reaches its stationary value \eqref{1.8} in about $30/(1+\beta)^2$ collisions per particle, while $X$ goes to zero in a significantly shorter period.

\begin{figure}
\includegraphics[width=8cm]{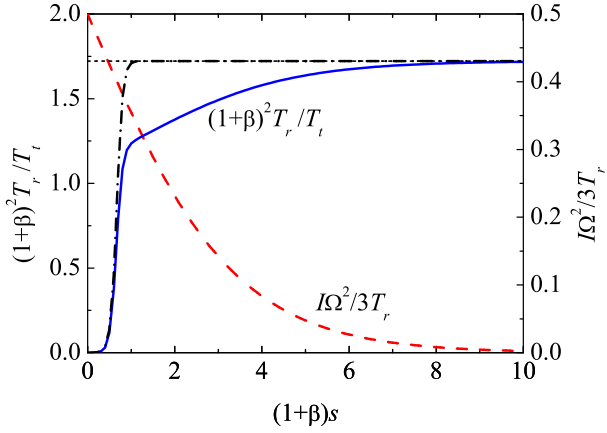}
  \caption{{(Color online) Plot of  $X=I\Omega^2/3\Tr$ (dashed line) and  $(1+\beta)^2\theta=(1+\beta)^2\Tr/\Tt$ (solid line) as functions of the number of collisions per particle ($s$) scaled by $(1+\beta)^{-1}$ for $\kappa=\frac{2}{5}$, $\alpha=0.8$, and $\beta=-0.99$,  starting from the initial condition $\theta(0)=1$, $X(0)=\frac{1}{2}$. The dash-dotted line represents $(1+\beta)^2\theta$ if $X(0)=0$. The horizontal line corresponds to the  stationary value $(1+\beta)^2\theta_\hcs$.}}
  \label{nfig2}
\end{figure}

{The special quasismooth limit $\beta\to -1$ deserves further comments \cite{S11b}. In that case, the asymptotic rotational-translational temperature ratio $\theta_\hcs$ diverges as $\theta_\hcs \to [(1+\kappa)^2/\kappa](1-\alpha^2)(1+\beta)^{-2}$, the two eigenvalues becoming $\ell_1\to \frac{5}{12}(1-\alpha^2)$, which is just the cooling rate of perfectly smooth spheres \cite{vNE98}, and $\ell_2\to \frac{5}{6}(1+\beta)/(1+\kappa)\to 0$. Therefore, if $X(0)=0$, $(1+\beta)^2\theta(s)$ relaxes to  $(1+\beta)^2\theta_\hcs$ after a \emph{finite} number of collisions per particle on the order of $\ell_1^{-1}$. On the other hand, if $X(0)\neq 0$, $(1+\beta)^2\theta(s)$ evolves in two well-defined stages. The first stage lasts a characteristic time $\sim \ell_1^{-1}$ and is very similar to that of the case $X(0)=0$. This first stage is followed by a much slower relaxation (with $s\sim \ell_2^{-1}\sim (1+\beta)^{-1}\to\infty$ collisions per particle) toward the asymptotic value. This singular scenario in the quasismooth limit is illustrated by Fig.\ \ref{nfig2} for $\beta=-0.99$ and $\alpha=0.8$.}

\begin{figure}
\includegraphics[width=8cm]{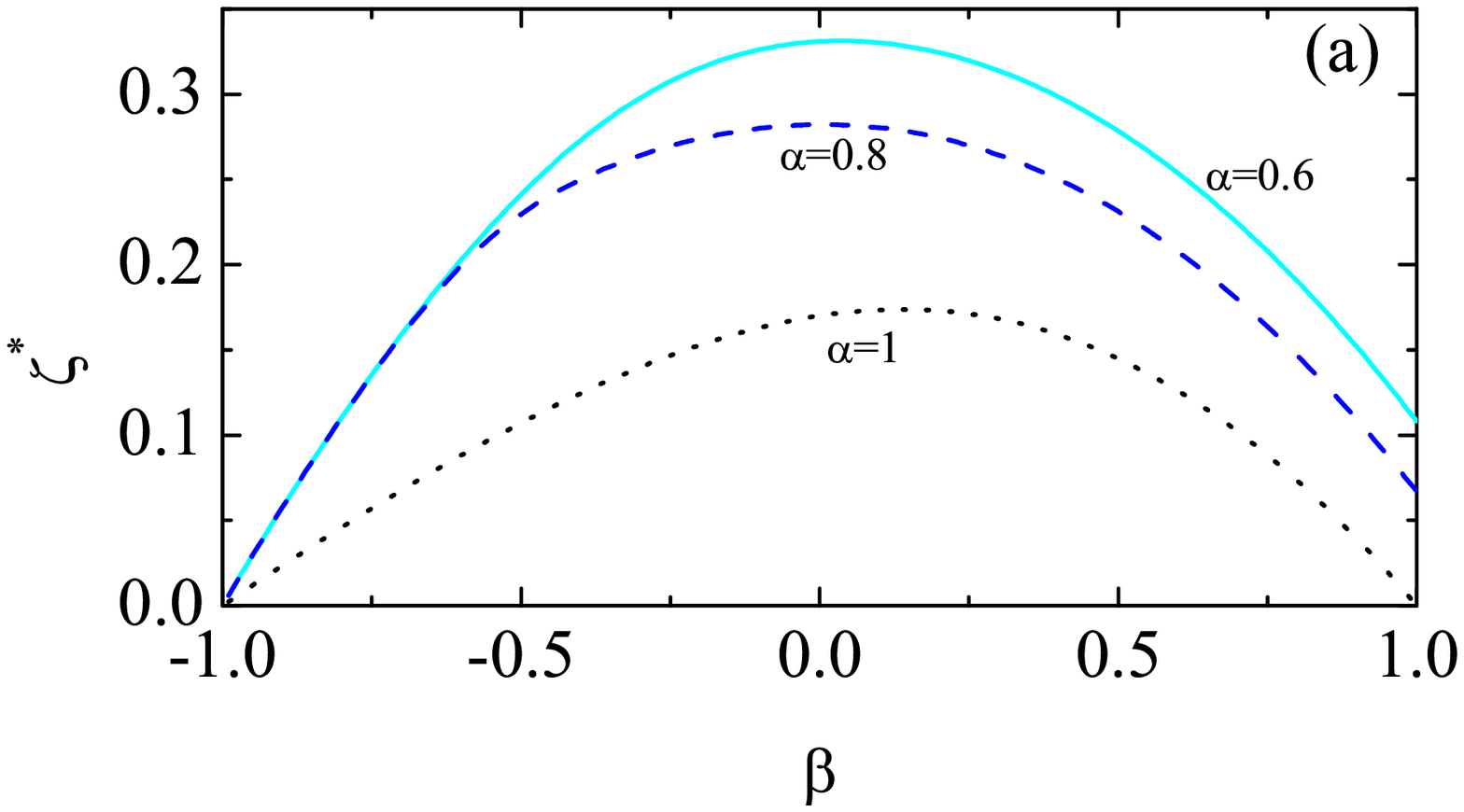}\\
\vskip0.2cm
\includegraphics[width=8cm]{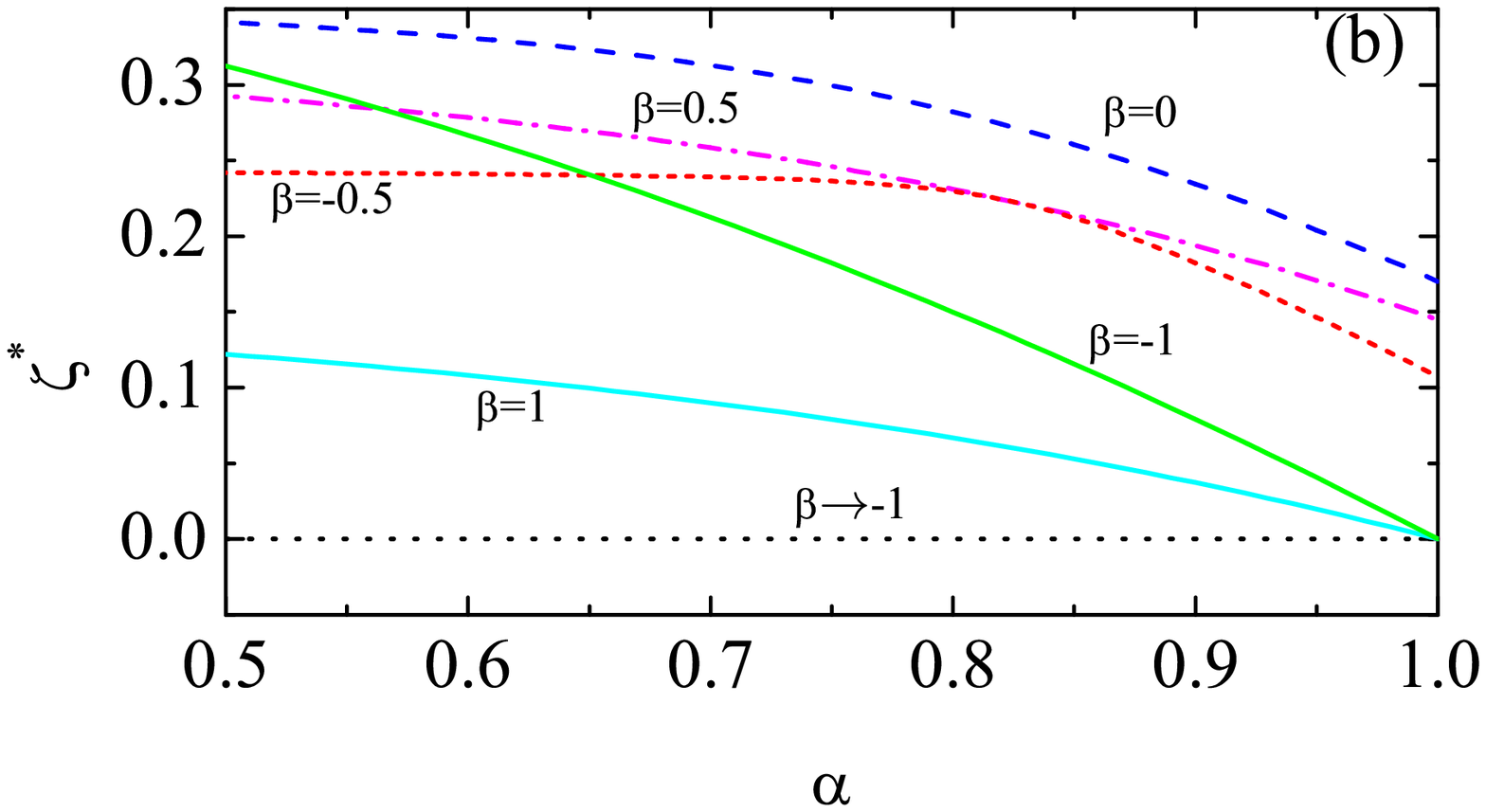}\\
\vskip0.2cm
\includegraphics[width=8cm]{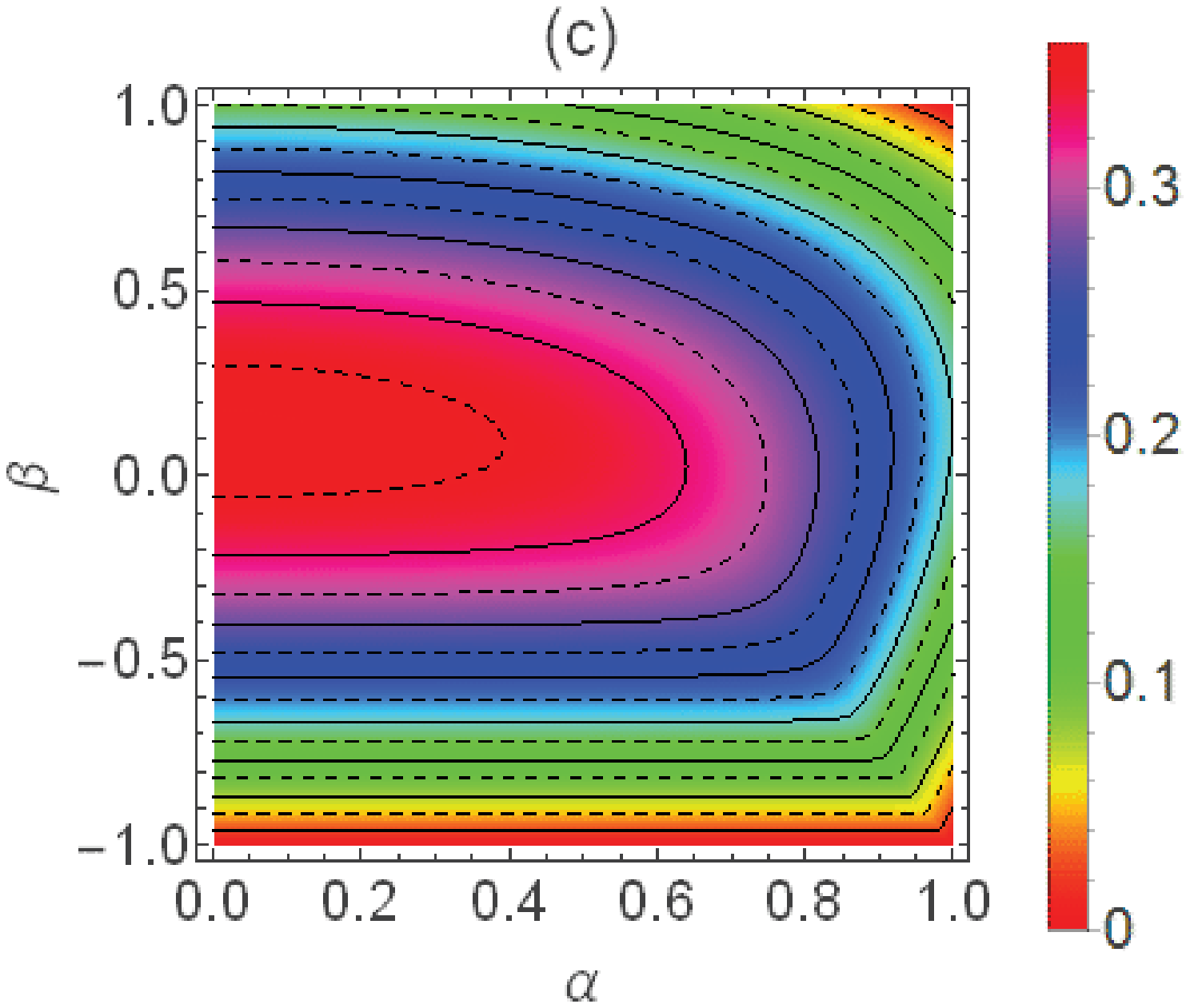}
\caption{{(Color online) (a) Reduced cooling rate $\zeta^*$ as a function of $\beta$ for $\kappa=\frac{2}{5}$ and $\alpha=0.6$, $0.8$, and $1$. (b) Reduced cooling rate $\zeta^*$ as a function of $\alpha$ for $\kappa=\frac{2}{5}$ and $\beta=-0.5$, $0$, $0.5$, and $1$, as well as in the quasismooth limit $\beta\to -1$ and for purely smooth spheres ($\beta=-1$). (c) Density plot of the reduced cooling rate $\zeta^*$ for $\kappa=\frac{2}{5}$. The contour lines correspond to $\zeta^*=0.025,0.05,\ldots,0.35$.}}
\label{nfig3}
\end{figure}

The stationary solution \eqref{1.8} represents the value of the temperature ratio in the HCS \cite{HZ97,LHMZ98,HHZ00,AHZ01,Z06}. In such a state the whole time dependence of the distribution function only occurs through the granular temperature $T(t)$, so that the Boltzmann equation \eqref{15a} becomes
\beq
-\zeta T \partial_T f =J[f,f],
\label{B_HCS}
\eeq
{where, according to Eq.\ \eqref{2.14}, $\zeta=\zeta^*\nu$ with}
\beq
{\zeta^*=\frac{5}{12}\frac{1}{1+\theta_\hcs}\left[1-\alpha^2+({1-\beta^2})
\frac{\theta_\hcs+\kappa}{1+\kappa}\right].}
\label{zetaHCS}
\eeq
{The dependence of the HCS (reduced) cooling rate $\zeta^*$ on both $\alpha$ and $\beta$ is displayed in Fig.\ \ref{nfig3}. As can be observed, at a given value of $\alpha$ a maximum of $\zeta^*$ occurs around $\beta\approx 0$. In the region close to $\beta=-1$, where $\theta_\hcs\gg 1$, Eq.\ \eqref{zetaHCS} becomes $\zeta^*\approx \frac{5}{12}(1-\beta^2)/(1+\kappa)$. Therefore, $\lim_{\beta\to -1}\zeta^*=0$.}

The HCS distribution function has the scaling form
\beq
f(\bc,\bw,t)=n\left(\frac{mI}{\tau_t\tau_r}\right)^{3/2} \left[T(t)\right]^{-3}\phi\left(\mathbf{c}(t),\mathbf{w}(t)\right),
\label{phi}
\eeq
where
\beq
\mathbf{c}(t)={\mathbf{V}\over\sqrt{2\tau_t T(t)/m}},\quad \mathbf{w}(t)={\bw\over\sqrt{2\tau_r T(t)/m}},
\label{cw}
\eeq
with the time-independent temperature ratios
\beq
\tau_t\equiv {\Tt(t)\over T(t)}={2\over1+\theta_\hcs},\quad \tau_r\equiv {\Tr(t)\over T(t)}={2\theta_\hcs\over1+\theta_\hcs}.
\eeq
Hence, according to Eq.\ \eqref{phi}, one has the relation
\beq
 T \partial_T f
=-\frac{1}{2}\left(\frac{\partial }{\partial\mathbf{V}}\cdot \mathbf{V} +\frac{\partial }{\partial\bw}\cdot \bw\right)f.
\eeq

{In the particular case of perfectly elastic and smooth particles ($\alpha=1$, $\beta= -1$), the rotational velocities must be ignored (i.e., $\tau_r=0$, $\tau_t=2$) and the solution of Eq.\ \eqref{B_HCS} is just the equilibrium distribution of translational velocities. In the other conservative case of perfectly elastic and rough spheres ($\alpha=\beta= 1$),  the solution of Eq.\ \eqref{B_HCS} is the equilibrium distribution with a common temperature (i.e., $\tau_r=\tau_t=1$).}
In the general dissipative case, however,  the  solution of Eq.\ \eqref{B_HCS} is not exactly known, although good approximations are given in the form of a two-temperature Maxwellian multiplied by  truncated Sonine polynomial expansions \cite{HZ97,ML98,LHMZ98,HHZ00,AHZ01,CLH02,Z06,BPKZ07,KBPZ09,BB12,SKS11,VSK14}.

\section{Chapman--Enskog method}
\label{sec4}

\subsection{Outline of the method}
To determine the distribution function from the Chapman--Enskog method \cite{CC70,FK72}, we write the Boltzmann equation (\ref{15a}) as
\be
\mathcal{D}_tf+\epsilon \mathbf{V}\cdot\bnabla f =J[f,f],
\ee{22}
where  $\epsilon$ is a uniformity parameter (set equal to unity at the end of the calculations) measuring the strength of the spatial gradients.
According to the method, the distribution function and the material time derivative are expanded in terms of the parameter $\epsilon$ as follows,
\be
f=f^{(0)}+\epsilon f^{(1)}+\epsilon^2 f^{(2)}+\cdots,
\ee{23a}
\be
\mathcal{D}_t=\mathcal{D}_t^{(0)}+\epsilon \mathcal{D}_t^{(1)}+\epsilon^2 \mathcal{D}_t^{(2)}+\cdots.
\ee{23b}
Substitution of Eq.\ \eqref{23a} into the definitions of the fluxes and the cooling rate gives
\beq
P_{ij}=p^{(0)}\delta_{ij}+\epsilon P_{ij}^{(1)}+\epsilon^2 P_{ij}^{(2)}+\cdots,
\label{P0}
\eeq
\beq
\mathbf{q}=\epsilon \mathbf{q}^{(1)}+\epsilon^2 \mathbf{q}^{(2)}+\cdots,
\eeq
\be
\zeta=\zeta^{(0)}+\epsilon \zeta^{(1)}+\epsilon^2 \zeta^{(2)}+\cdots,
\ee{23c}
where $p^{(0)}=n\tau_t  T$,
\beq
\zeta^{(0)}=-\frac{1}{6nT}\int d\bc\int d\bw\,\left({mV^2}+{I\omega^2}\right)J[f^{(0)},f^{(0)}],
\eeq
\bal
\label{zeta1}
\zeta^{(1)}=&\frac{1}{6nT}\int d\bc\int d\bw\,\left({mV^2}+{I\omega^2}\right)\mathcal{L}f^{(1)}\nn
=&\frac{\tau_t}{2}\zeta_t^{(1)}+\frac{\tau_r}{2}\zeta_r^{(1)}.
\eal
In Eq.\ \eqref{zeta1} $\mathcal{L}$ is the linearized collision operator defined as
\beq
\mathcal{L}\Phi=-J[\Phi,f^{(0)}]-J[f^{(0)},\Phi].
\eeq
Thus, $\zeta^{(1)}=\mathcal{J}_L[mV^2+I\omega^2|f^{(1)}]/6nT$, with the notation
\bal
\label{16L}
\mathcal{J}_L[\psi|\Phi]\equiv&\int d\bc\int d\bw\,\psi(\bx,\bc,\bw,t)\mathcal{L}\Phi\nn
=& {\d^2\over2}\int d\bc\int d\bw\int d\bc_1\int d\bw_1\int_+ d\bk\,(\bk\cdot\bg)\nn
&\times\left(\psi_1+\psi-\psi_1'-\psi'\right)(f_1^{(0)}\Phi+f^{(0)}\Phi_1),
\eal
where $\psi_1\equiv \psi(\bc_1,\bw_1)$, $\psi'\equiv \psi(\bc',\bw')$, $\psi_1'\equiv \psi(\bc_1',\bw_1')$, $f_1^{(0)}\equiv f^{(0)}(\bc_1,\bw_1)$, $\Phi_1\equiv \Phi(\bc_1,\bw_1)$, and in the last step use has been made of Eq.\ \eqref{16J}.

Insertion of the expansions (\ref{23a}) and \eqref{23b} into the Boltzmann equation (\ref{22}) leads to the corresponding integro-differential equations for the different orders $f^{(k)}$. In particular, the two first equations are
\beq\label{24a}
  {\cal D}_t^{(0)} f^{(0)}=J[f^{(0)},f^{(0)}],
  \eeq
\beq
 \label{24b}
 \left({\cal D}_t^{(0)} +\mathcal{L}\right)f^{(1)}=-\left({\cal D}_t^{(1)}+\mathbf{V}\cdot\bnabla \right)f^{(0)}.
 \eeq

Since the distribution function in \eqref{23a} depends on time and space only through its dependence on the hydrodynamic fields $n$, $\buu$, and $T$,
the action of the operator $\mathcal{D}_t^{(k)}$ can be written as
\beq
\mathcal{D}_t^{(k)}=\left(\mathcal{D}_t^{(k)} n\right)\frac{\partial}{\partial n}+ \left(\mathcal{D}_t^{(k)} \buu\right)\cdot\frac{\partial}{\partial \buu}+\left(\mathcal{D}_t^{(k)} T\right)\frac{\partial}{\partial T}.
\label{Dtk}
\eeq
{}From the balance equations \eqref{18}--\eqref{20} it follows that
\beq\label{26a}
{\cal D}_t^{(0)}n=0,\quad {\cal D}_t^{(1)}n=-n\bnabla\cdot\buu,
\eeq
\beq
\label{26b}
{\cal D}_t^{(0)}\buu=0,\quad {\cal D}_t^{(1)}\buu=-{\tau_t\over\rho}\left(n\bnabla T+T\bnabla n\right),
\eeq
\beq
\lb{26c}
{\cal D}_t^{(0)} T =- T  \zeta^{(0)},\quad{\cal D}_t^{(1)} T =- T  \zeta^{(1)}-{ \tau_t\over3}T\bnabla\cdot\buu.
\eeq

Taking into account that ${\cal D}_t^{(0)}\Phi=-\zeta^{(0)}T\partial_T\Phi$, it is obvious that Eq.\ \eqref{24a} is formally equivalent to the HCS equation \eqref{B_HCS}. This means that $f^{(0)}$ is the \emph{local} version of the HCS distribution. Moreover, in the approximation \eqref{3.1}, {$\zeta^*=\zeta^{(0)}/\nu$ is given by Eq.\ \eqref{zetaHCS}}.

\subsection{First-order distribution}

The first-order function $f^{(1)}$ obeys the linear equation \eqref{24b}. By using the properties \eqref{Dtk}--\eqref{26c}, the inhomogeneous term of Eq.\ \eqref{24b} becomes
\bal
-\left({\cal D}_t^{(1)}+\mathbf{V}\cdot\bnabla\right)f^{(0)}=&\mathbf{A}\cdot\bnabla \ln T+\mathbf{B}\cdot\bnabla \ln n{+C_{ij}\nabla_j u_i}\nn
&+E\bnabla\cdot\buu+T\zeta^{(1)}\partial_T f^{(0)},
\label{ABCE}
\eal
where
\beq
\mathbf{A}=-T\left(\mathbf{V}\partial_T+\frac{\tau_t}{m}\partial_{\mathbf{V}}\right)f^{(0)},
\label{eqA}
\eeq
\beq
\mathbf{B}=-\left(\mathbf{V}+\frac{\tau_t T}{m}\partial_{\mathbf{V}}\right)f^{(0)},
\label{eqB}
\eeq
\beq
{C_{ij}=\left(\partial_{V_i}V_j-\frac{1}{3}\delta_{ij}\partial_{\mathbf{V}}\cdot\mathbf{V}\right)f^{(0)}},
\label{eqC}
\eeq
\beq
E=\frac{1}{3}\left(\partial_\mathbf{V}\cdot\mathbf{V}+\tau_tT\partial_T\right)f^{(0)}.
\label{eqE}
\eeq

In the case of pure smooth particles, $f^{(0)}$ is a function of the translational velocity only, there is no rotational energy  and hence $\tau_r=0$, $\tau_t=2$, and $T\partial_T f^{(0)}=-\frac{1}{2}\partial_{\mathbf{V}}\cdot\mathbf{V} f^{(0)}$. As a consequence, $E=0$ and one recovers the known results for inelastic smooth  particles \cite{BDKS98}. Thus, the presence of roughness induces a non-vanishing function $E$, even in the conservative case of perfectly rough particles ($\alpha=\beta=1$) \cite{CC70,K10a}. A subtler consequence of roughness is the symmetry breakdown of the traceless tensor $C_{ij}$. Isotropy implies that $f^{(0)}(\mathbf{V},\bw)$ is a function of the three scalars $V^2$, $\omega^2$ and $\chi^2\equiv(\mathbf{V}\cdot\bw)^2$. Therefore,
\beq
{C_{ij}- C_{ji}=2\frac{\partial f^{(0)}}{\partial \chi^2}(\mathbf{V}\cdot\bw)(V_j\omega_i-V_i\omega_j)}.
\eeq
If one neglects the orientational correlations between $\mathbf{V}$ and $\bw$ in the HCS so that the dependence of $f^{(0)}$ on $\chi^2$ is ignored, then $C_{ij}=C_{ji}$, as happens in the pure smooth case.

Taking into account Eq.\ \eqref{ABCE}, the solution to Eq.\ \eqref{24b} has the form
\beq
f^{(1)}=
\bbA\cdot\bnabla \ln T+\bbB\cdot\bnabla \ln n+\bbC_{ij}{\nabla_j u_i}
+\bbE\bnabla\cdot\buu,
\label{f1}
\eeq
where the vectors $\bbA$ and $\bbB$, the traceless tensor $\bbC_{ij}$, and the scalar $\bbE$ are unknown functions to be determined.
Combination of Eqs.\ \eqref{zeta1} and \eqref{f1} allows us to write the first-order contribution to the cooling rate as
\beq
\zeta^{(1)}={-\xi}\bnabla\cdot\buu
\label{zeta1xi}
\eeq
with
\beq
\xi={\frac{1}{2}\left(\tau_t\xi_t+\tau_r\xi_r\right)},
\label{xi}
\eeq
where
\beq
{\xi_t=-\frac{m}{3n\tau_tT}\mathcal{J}_L[V^2|\bbE],\quad \xi_r=-\frac{I}{3n\tau_rT}\mathcal{J}_L[\omega^2|\bbE]}
\label{xitr}
\eeq
are the translational and rotational contributions to $\xi$.

Substitution of Eq.\ \eqref{f1} into Eq.\ \eqref{24b} gives the following set of linear integral equations:
\beq
\left(-\frac{\zeta^{(0)}}{2}-\zeta^{(0)}T\partial_T+\mathcal{L}\right)\bbA=\mathbf{A},
\label{eqLA}
\eeq
\beq
\left(-\zeta^{(0)}T\partial_T+\mathcal{L}\right)\bbB=\mathbf{B}+\zeta^{(0)}\bbA,
\label{eqLB}
\eeq
\beq
\left(-\zeta^{(0)}T\partial_T+\mathcal{L}\right)\bbC_{ij}=C_{ij},
\label{eqLC}
\eeq
\beq
\left(-\zeta^{(0)}T\partial_T+\mathcal{L}\right)\bbE{+\xi} T\partial_T f^{(0)}=E.
\label{eqLE}
\eeq
In Eqs.\ \eqref{eqLA} and \eqref{eqLB} use has been made of the property
\bal
\mathcal{D}_t^{(0)}\bnabla \ln T=&\bnabla \mathcal{D}_t^{(0)}\ln T=-\bnabla\zeta^{(0)}\nn
=&-\zeta^{(0)}\left(\bnabla \ln n+\frac{1}{2}\bnabla \ln T\right).
\eal

According to Eqs.\ \eqref{eqA}--\eqref{eqE}, the functions $\mathbf{A}$, $\mathbf{B}$, $C_{ij}$ and $E$ are orthogonal to $\{1,\mathbf{V}, mV^2+I\omega^2,\bw\}$, i.e.,
\beq
\int d\bc\int d\bw\, \left(
\begin{array}{c}
  1\\
  \mathbf{V}\\
  mV^2+I\omega^2\\
  \bw
  \end{array}
  \right)
  \cdot
  \left(\mathbf{A},\mathbf{B},C_{ij},E\right)=0.
  \eeq
Therefore, the Fredholm alternative \cite{DS67} implies that the necessary conditions for the existence of solutions (solubility conditions) to Eqs.\ \eqref{eqLA}--\eqref{eqLE}  are
\beq
\int d\bc\int d\bw\, \left(
\begin{array}{c}
  1\\
  \mathbf{V}\\
  mV^2+I\omega^2\\
  \bw
  \end{array}
  \right)
  \cdot
  \left(\bbA,\bbB,\bbC_{ij},\bbE\right)=0.
  \label{solub}
  \eeq
The solubility conditions associated with $\{1,\mathbf{V}, mV^2+I\omega^2\}$ mean that, by construction, the hydrodynamic quantities $n$, $\mathbf{u}$, and $T$ are fully contained in $f^{(0)}$. As for the solubility condition associated with $\bw$, it implies
\beq
\boldsymbol{\Omega}^{(1)}=\frac{1}{n}\int d\bc\int d\bw\, \bw f^{(1)}=0.
\eeq
  \section{Navier--Stokes--Fourier transport coefficients}
  \label{sec5}
\subsection{Exact formal expressions}

{}From Eq.\ \eqref{f1} one can express the first-order pressure tensor and heat flux as
\beq
P_{ij}^{(1)}=-\eta \left(\nabla_i u_j+\nabla_j u_i-\frac{2}{3}\delta_{ij}\bnabla\cdot\buu \right)-\eta_b\delta_{ij}\bnabla\cdot \buu,
\label{Pij1}
\eeq
\beq
\mathbf{q}^{(1)}=-\lambda \bnabla T-\mu\bnabla n,
\label{q1}
\eeq
where the transport coefficients are
\beq
\eta=-\frac{m}{10}\int d\bc\int d\bw\, \left(V_iV_j-\frac{V^2}{3}\delta_{ij}\right)\bbC_{ij},
\eeq
\beq
\eta_b=-\frac{m}{3}\int d\bc\int d\bw\, V^2\bbE,
\eeq
\beq
\lambda={\tau_t\lambda_t+\tau_r\lambda_r},\quad \mu={\mu_t+\mu_r},
\eeq
with
\beq
\lambda_t=-\frac{m}{6\tau_tT}\int d\bc\int d\bw\, V^2\mathbf{V}\cdot\bbA,
\eeq
\beq
 \lambda_r=-\frac{I}{6\tau_rT}\int d\bc\int d\bw\, \omega^2\mathbf{V}\cdot\bbA,
\eeq
\beq
\mu_t=-\frac{m}{6n}\int d\bc\int d\bw\, V^2\mathbf{V}\cdot\bbB,
\eeq
\beq
\mu_r=-\frac{I}{6n}\int d\bc\int d\bw\, \omega^2\mathbf{V}\cdot\bbB.
\eeq
In the constitutive equations \eqref{Pij1} and \eqref{q1}, $\eta$ is the shear viscosity, $\eta_b$ is the bulk viscosity, $\lambda$ is the thermal conductivity, and $\mu$ is a Dufour-like coefficient. The two latter coefficients have translational ($\lambda_t$, $\mu_t$) and rotational ($\lambda_r$, $\mu_r$) contributions.

By multiplying Eq.\ \eqref{eqLA} by  $V^2\mathbf{V}$ and $\omega^2\mathbf{V}$, and integrating over velocity one obtains
\beq
{\lambda_t=\frac{5n\tau_t T}{2m}\frac{1+2 a_{20}^{(0)}}{\nu_{\lambda_t}-2\zeta^{(0)}}},
\label{lambdat}
\eeq
\beq
{\lambda_r=\frac{3n\tau_t T}{2m}\frac{1+2a_{11}^{(0)}}{\nu_{\lambda_r}-2\zeta^{(0)}}},
\label{lambdar}
\eeq
where
\beq
{\nu_{\lambda_t}=\frac{\int d\bc\int d\bw\, V^2\mathbf{V}\cdot\mathcal{L}\bbA}{\int d\bc\int d\bw\, V^2\mathbf{V}\cdot\bbA}},
\label{nulambdat}
\eeq
\beq
{\nu_{\lambda_r}=\frac{\int d\bc\int d\bw\, \omega^2\mathbf{V}\cdot\mathcal{L}\bbA}{\int d\bc\int d\bw\, \omega^2\mathbf{V}\cdot\bbA}}
\label{nulambdar}
\eeq
are the associated collision frequencies and
\beq
{a_{20}^{(0)}=\frac{m^2}{15n\tau_t^2 T^2}\int d\bc\int d\bw\, V^4 f^{(0)}-1},
\eeq
\beq
{a_{11}^{(0)}=\frac{m I}{9n\tau_t\tau_r T^2}\int d\bc\int d\bw\, V^2\omega^2 f^{(0)}-1},
\eeq
are cumulants of the HCS distribution $f^{(0)}$ \cite{VSK14}. Upon deriving Eqs.\ \eqref{lambdat} and \eqref{lambdar} we have taken into account that, by dimensional analysis, $T\partial_T\lambda_t=\frac{1}{2}\lambda_t$ and $T\partial_T\lambda_r=\frac{1}{2}\lambda_r$. Analogously, from Eq.\ \eqref{eqLB} one gets
\beq
{\mu_t=\frac{\tau_t T}{n}\frac{\lambda_t\zeta^{(0)}+(5n\tau_t T/2m)a_{20}^{(0)}}{\nu_{\mu_t}-\frac{3}{2}\zeta^{(0)}}},
\label{mut}
\eeq
\beq
{\mu_r=\frac{\tau_r T}{n}\frac{\lambda_r\zeta^{(0)}+(3n\tau_t T/2m)a_{11}^{(0)}}{\nu_{\mu_r}-\frac{3}{2}\zeta^{(0)}}},
\label{mur}
\eeq
where
\beq
{\nu_{\mu_t}=\frac{\int d\bc\int d\bw\, V^2\mathbf{V}\cdot\mathcal{L}\bbB}{\int d\bc\int d\bw\, V^2\mathbf{V}\cdot\bbB}},
\label{numut}
\eeq
\beq
{\nu_{\mu_r}=\frac{\int d\bc\int d\bw\, \omega^2\mathbf{V}\cdot\mathcal{L}\bbB}{\int d\bc\int d\bw\, \omega^2\mathbf{V}\cdot\bbB}},
\label{numur}
\eeq
and use has been made of $T\partial_T\mu_t=\frac{3}{2}\mu_t$ and $T\partial_T\mu_r=\frac{3}{2}\mu_r$.

Next, multiplication of Eq.\ \eqref{eqC} by $V_iV_j-\frac{1}{3}\delta_{ij}V^2$ and integration over velocity yields
\beq
\eta=\frac{n\tau_t T}{\nu_\eta-\frac{1}{2}\zeta^{(0)}},
\label{eta}
\eeq
where
\beq
\nu_\eta=\frac{\int d\bc\int d\bw\, \left(V_iV_j-\frac{V^2}{3}\delta_{ij}\right)\mathcal{L}\bbC_{ij}}{\int d\bc\int d\bw\, \left(V_iV_j-\frac{V^2}{3}\delta_{ij}\right)\bbC_{ij}}.
\label{nueta}
\eeq
Finally, multiplying Eq.\ \eqref{eqLE} by $V^2$ allows us to obtain
\beq
\eta_b=\frac{\tau_t\tau_r nT}{\zeta^{(0)}}{\left(\xi_t-\xi_r-\frac{2}{3}\right)}.
\label{etab}
\eeq
In Eqs.\ \eqref{eta} and \eqref{etab} we have made use of the properties $T\partial_T\eta=\frac{1}{2}\eta$ and $T\partial_T\eta_b=\frac{1}{2}\eta_b$.

The existence of a nonzero bulk viscosity induces a breakdown of energy equipartition additional to the one already present in the HCS.
Taking the trace in Eqs.\ \eqref{P0} and \eqref{Pij1}, one has
\beq
\Tt=\tau_t T-\frac{\eta_b}{n}\bnabla\cdot\mathbf{u}+\cdots,
\eeq
where the ellipses denote terms of at least second order in the hydrodynamic gradients. Since the total temperature $T$ is not affected by the gradients, then
\beq
 \Tr=\tau_r T+\frac{\eta_b}{n}\bnabla\cdot\mathbf{u}+\cdots.
\eeq
Thus the temperature ratio becomes
\beq
\frac{\Tr}{\Tt}=\frac{\tau_r}{\tau_t}\left[1+2\frac{\xi_t-\xi_r-\frac{2}{3}}{\zeta^{(0)}}\bnabla\cdot\mathbf{u}+\cdots\right],
\eeq
where use has been made of Eq.\ \eqref{etab}.

Equations \eqref{lambdat}, \eqref{lambdar}, \eqref{mut}, \eqref{mur}, \eqref{eta}, and \eqref{etab} are formally exact but require the solution of the set of linear integral equations \eqref{eqLA}--\eqref{eqLE}. As happens in the conservative case \cite{CC70,FK72}, the exact solution of those equations is not known. Since $\bbA$ is a vector, it can be expressed as a sum of projections along the three \emph{polar} vectors $\mathbf{V}$, $(\mathbf{V}\cdot\bw)\bw$, and $\mathbf{V}\times\bw$ \cite{CLD65}, namely
\beq
\bbA=\mathcal{A}_1 \mathbf{V}+\mathcal{A}_2(\mathbf{V}\cdot\bw)\bw+\mathcal{A}_3\mathbf{V}\times\bw,
\eeq
where $\mathcal{A}_i$ are unknown isotropic scalar functions, i.e., they depend on $V^2$, $\omega^2$, and $\chi^2=(\mathbf{V}\cdot\bw)^2$ only.
Of course, the vector function $\bbB$ has a similar structure. The tensor $\bbC_{ij}$ can be expressed as a combination of traceless dyadic products of the three vectors $\mathbf{V}$, $(\mathbf{V}\cdot\bw)\bw$, and $\mathbf{V}\times\bw$ with unknown scalar coefficients. Finally, $\bbE$ is an unknown scalar function.

In Sec.\ \ref{sec5B} we derive explicit expressions for the transport coefficients by considering the leading terms in a Sonine polynomial expansion.

\subsection{Sonine approximation\label{sec5B}}

As said in Sec.\ \ref{sec3}, the HCS distribution function $f^{(0)}$ is not exactly known. In a recent paper \cite{VSK14}, the first four relevant cumulants (in particular, $a_{20}^{(0)}$ and $a_{11}^{(0)}$) have been studied theoretically by means of a Sonine polynomial expansion and also computationally by means of the direct simulation Monte Carlo (DSMC) method \cite{B94} and event-driven molecular dynamics. The results show that three of the cumulants are in general relatively small. On the other hand, the cumulant
\beq
{a_{02}^{(0)}=\frac{I^2}{15n\tau_r^2 T^2}\int d\bc\int d\bw\, \omega^4 f^{(0)}-1}
\eeq
measuring the kurtosis of the angular velocity distribution can take values larger than unity in the region of small roughness. Outside that region, $a_{02}^{(0)}$  decreases as roughness increases. As an example (see Fig.\ 3 of Ref.\ \cite{VSK14}), at $\alpha=0.9$, $|a_{20}^{(0)}|\lesssim 0.02$ and $|a_{11}^{(0)}|\lesssim 0.07$ for  the whole range $-1\leq \beta\leq 1$, and $a_{02}^{(0)}\lesssim 0.2$ in the range $-0.5\lesssim \beta\leq 1$.

{}From a practical point of view, it must be noted that most of the
materials are characterized by positive values of the roughness
parameter (typically, $\beta\sim 0.5$) \cite{L99} and in those cases the cumulants are small. As a consequence, the HCS distribution can be rather well approximated by the \emph{two-temperature} Maxwellian
\beq
f^{(0)}\to f^{(0)}_M=n\Bigg({m I\over 4\pi^2\tau_t\tau_r T^2}\Bigg)^{3\over2}e^{-c^2-w^2},
\label{Maxw}
\eeq
where we recall that the scaled translational and angular velocities $\mathbf{c}$ and $\mathbf{w}$ are defined by Eq.\ \eqref{cw}. In this Maxwellian approximation, the functions \eqref{eqA}--\eqref{eqE} reduce to
\beq
\mathbf{A}\to- f^{(0)}_Mv_0\left(c^2+w^2-4\right)\mathbf{c},
\label{A_M}
\eeq
\beq
\mathbf{B}\to 0,
\eeq
\beq
C_{ij}\to -2f^{(0)}_M\left(c_ic_j-\frac{1}{3}c^2\delta_{ij}\right),
\eeq
\beq
E\to\frac{1}{3}f^{(0)}_M\left[\tau_t\left(w^2-\frac{3}{2}\right)-\tau_r\left(c^2-\frac{3}{2}\right)\right],
\label{E_M}
\eeq
where
\beq
v_0\equiv \sqrt{2\tau_t T/m}
\eeq
is the (translational) thermal speed.

The Maxwellian forms \eqref{A_M}--\eqref{E_M} suggest to approximate the unknown functions $\bbA$, $\bbB$, $\bbC_{ij}$, and $\bbE$ by
\beq
\bbA\to -f^{(0)}_M \frac{v_0}{\nu}\left[\gamma_{A_t}\left(c^2-\frac{5}{2}\right)+\gamma_{A_r}\left(w^2-\frac{3}{2}\right)\right]\mathbf{c},
\label{Aapprox}
\eeq
\beq
\bbB\to  -f^{(0)}_M\frac{v_0}{\nu}\left[\gamma_{B_t}\left(c^2-\frac{5}{2}\right)+\gamma_{B_r}\left(w^2-\frac{3}{2}\right)\right]\mathbf{c},
\eeq
\beq
\bbC_{ij}\to -f^{(0)}_M\frac{\gamma_C}{\nu} \left(c_ic_j-\frac{1}{3}c^2\delta_{ij}\right),
\eeq
\beq
\bbE\to f^{(0)}_M\frac{\gamma_E}{\nu}\left[\tau_t\left(w^2-\frac{3}{2}\right)-\tau_r\left(c^2-\frac{3}{2}\right)\right],
\label{Eapprox}
\eeq
where $\nu$ is defined by Eq.\ \eqref{nu} (with $\Tt=\tau_t T$) and the   $\gamma$ coefficients are directly related to the transport coefficients by
\beq
\eta=\frac{n\tau_t T}{\nu}\frac{\gamma_C}{2},\quad
\eta_b=\frac{n\tau_t\tau_r T}{\nu}\gamma_E,
\label{X1}
\eeq
\beq
\lambda_t=\frac{5}{2}\frac{n\tau_t T}{m\nu}\gamma_{A_t},\quad \lambda_r=\frac{3}{2}\frac{n\tau_t T}{m\nu}\gamma_{A_r},
\label{X2}
\eeq
\beq
\mu_t=\frac{5}{2}\frac{\tau_t^2 T^2}{m\nu}\gamma_{B_t},\quad \mu_r=\frac{3}{2}\frac{\tau_t \tau_r T^2}{m\nu}\gamma_{B_r}.
\label{X3}
\eeq
It can be checked that the forms \eqref{Aapprox}--\eqref{Eapprox} are consistent with the solubility conditions \eqref{solub}.

The basic Sonine approximations \eqref{Aapprox}--\eqref{Eapprox} allow us to evaluate explicitly the transport coefficients. The main steps are described in the Appendix and the final results are displayed in Table \ref{table1} \cite{note_14_05_2}.

\begin{figure}
\includegraphics[width=8cm]{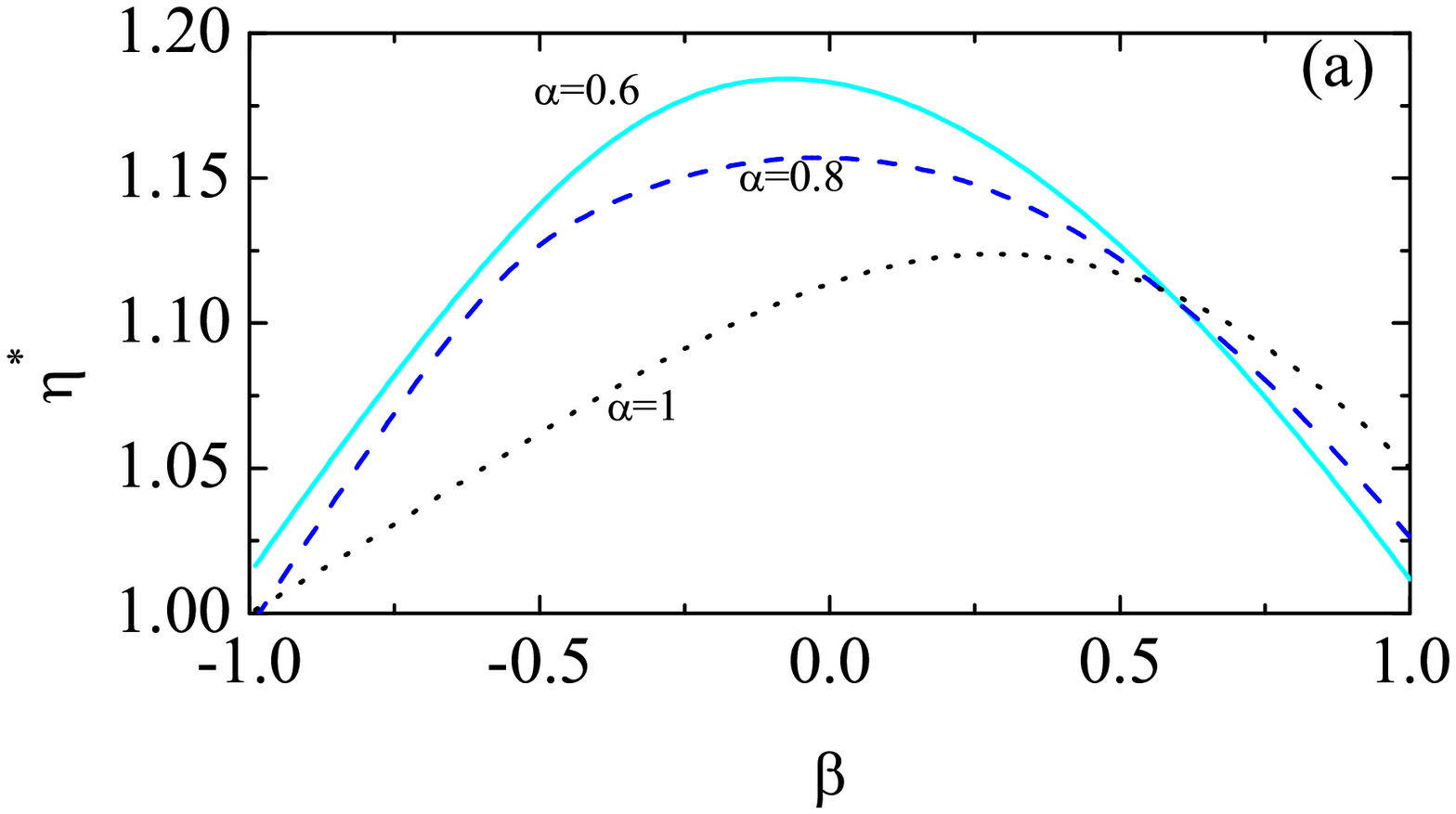}\\
\vskip0.2cm
\includegraphics[width=8cm]{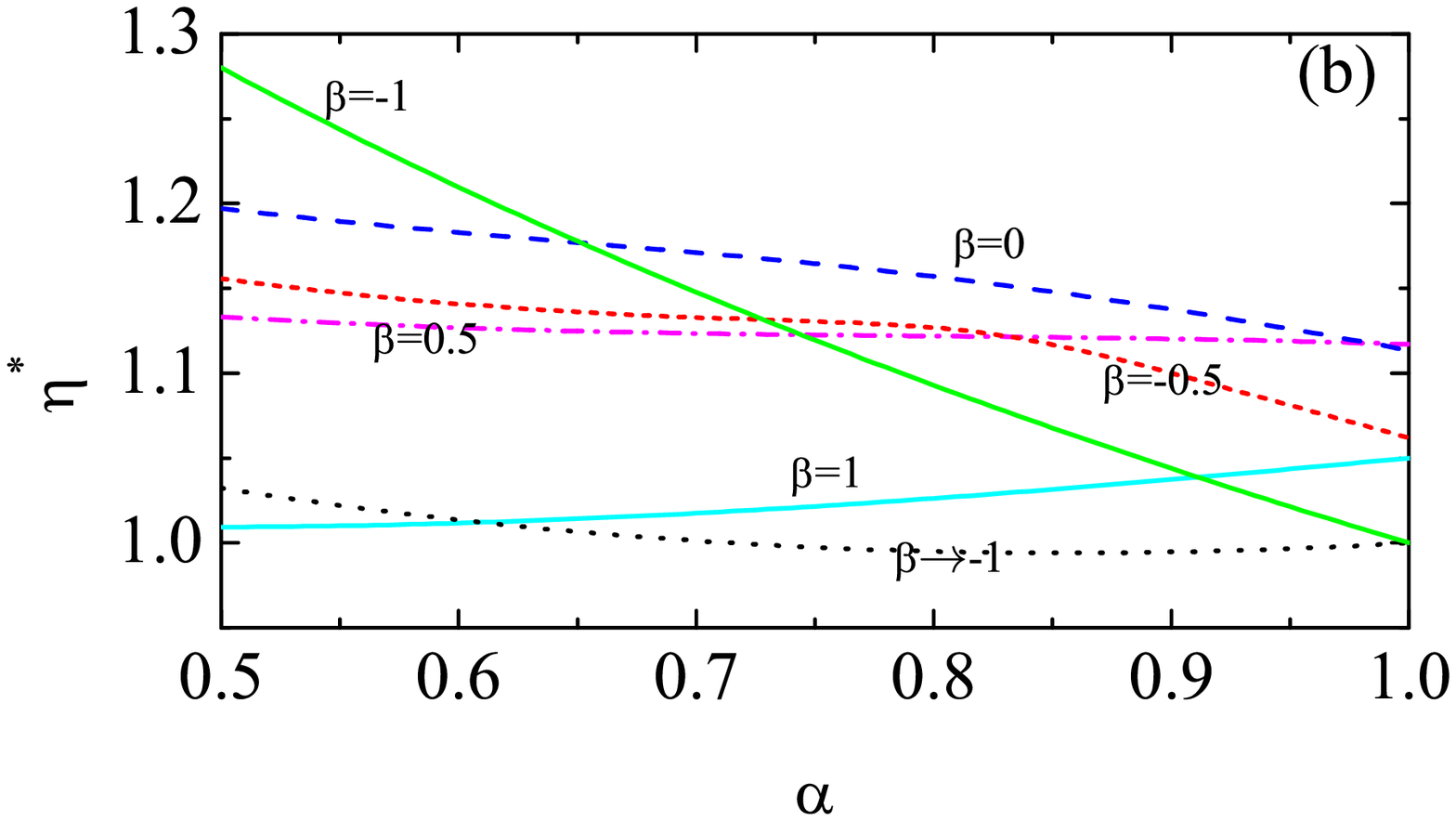}\\
\vskip0.2cm
\includegraphics[width=8cm]{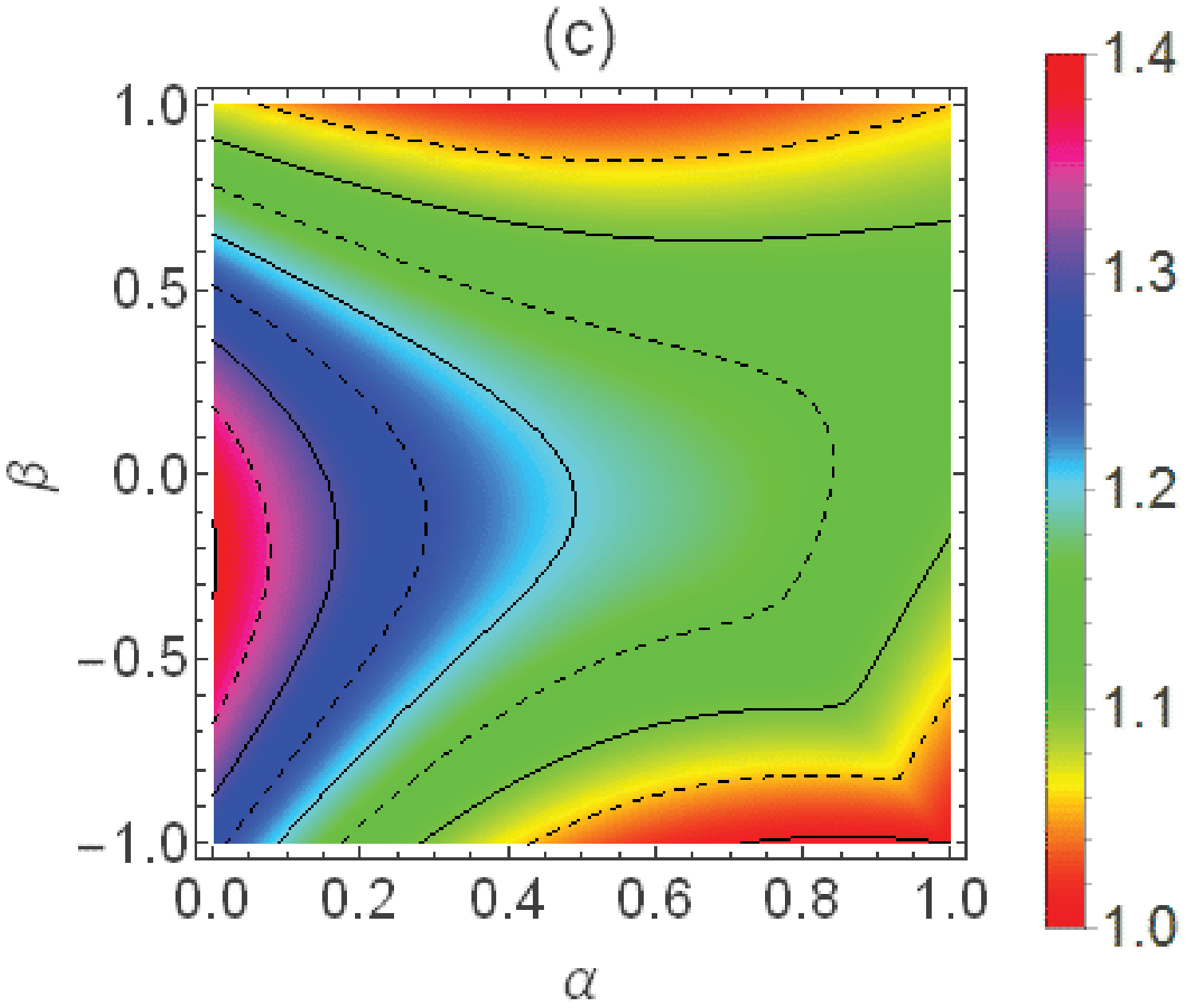}
\caption{(Color online) (a) Reduced shear viscosity $\eta^*$ as a function of $\beta$ for $\kappa=\frac{2}{5}$ and $\alpha=0.6$, $0.8$, and $1$. (b) Reduced shear viscosity $\eta^*$ as a function of $\alpha$ for $\kappa=\frac{2}{5}$ and $\beta=-0.5$, $0$, $0.5$, and $1$, as well as in the quasismooth limit $\beta\to -1$ and for purely smooth spheres ($\beta=-1$). (c) Density plot of the reduced shear viscosity $\eta^*$ for $\kappa=\frac{2}{5}$. The contour lines correspond to $\eta^*=1,1.05,\ldots,1.4$.}
\label{fig2}
\end{figure}

\begin{figure}
\includegraphics[width=8cm]{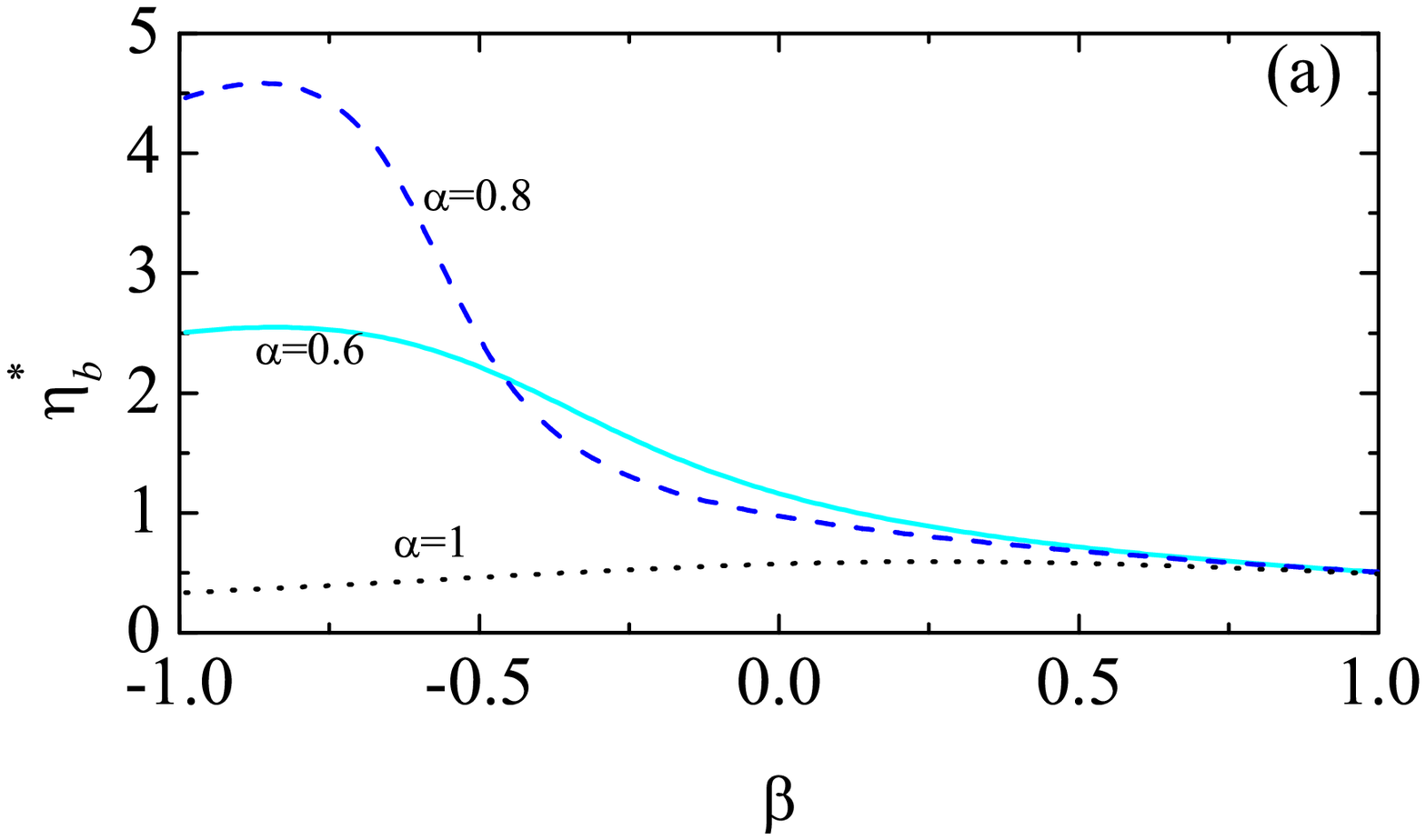}\\
\vskip0.2cm
\includegraphics[width=8cm]{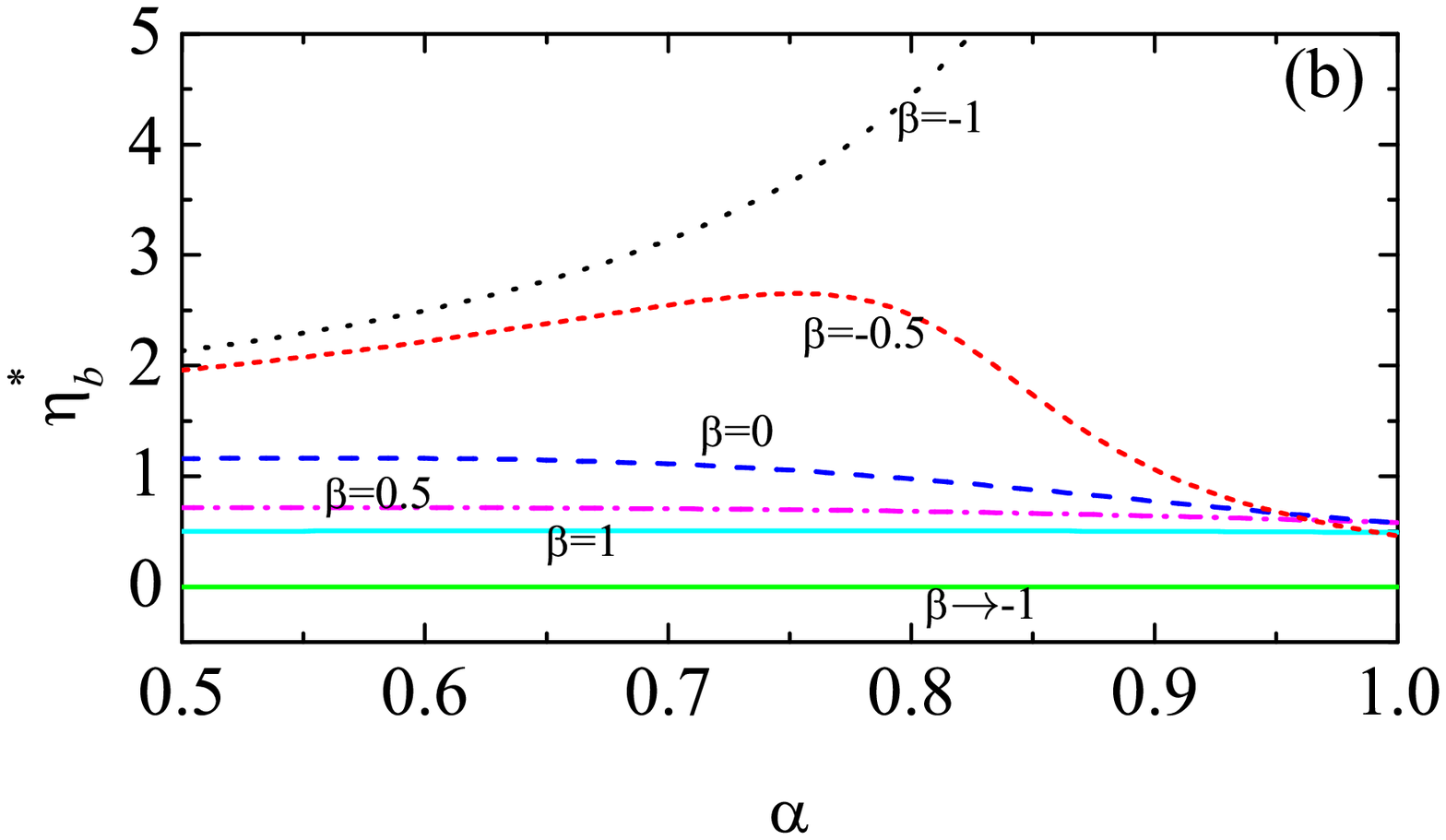}\\
\vskip0.2cm
\includegraphics[width=8cm]{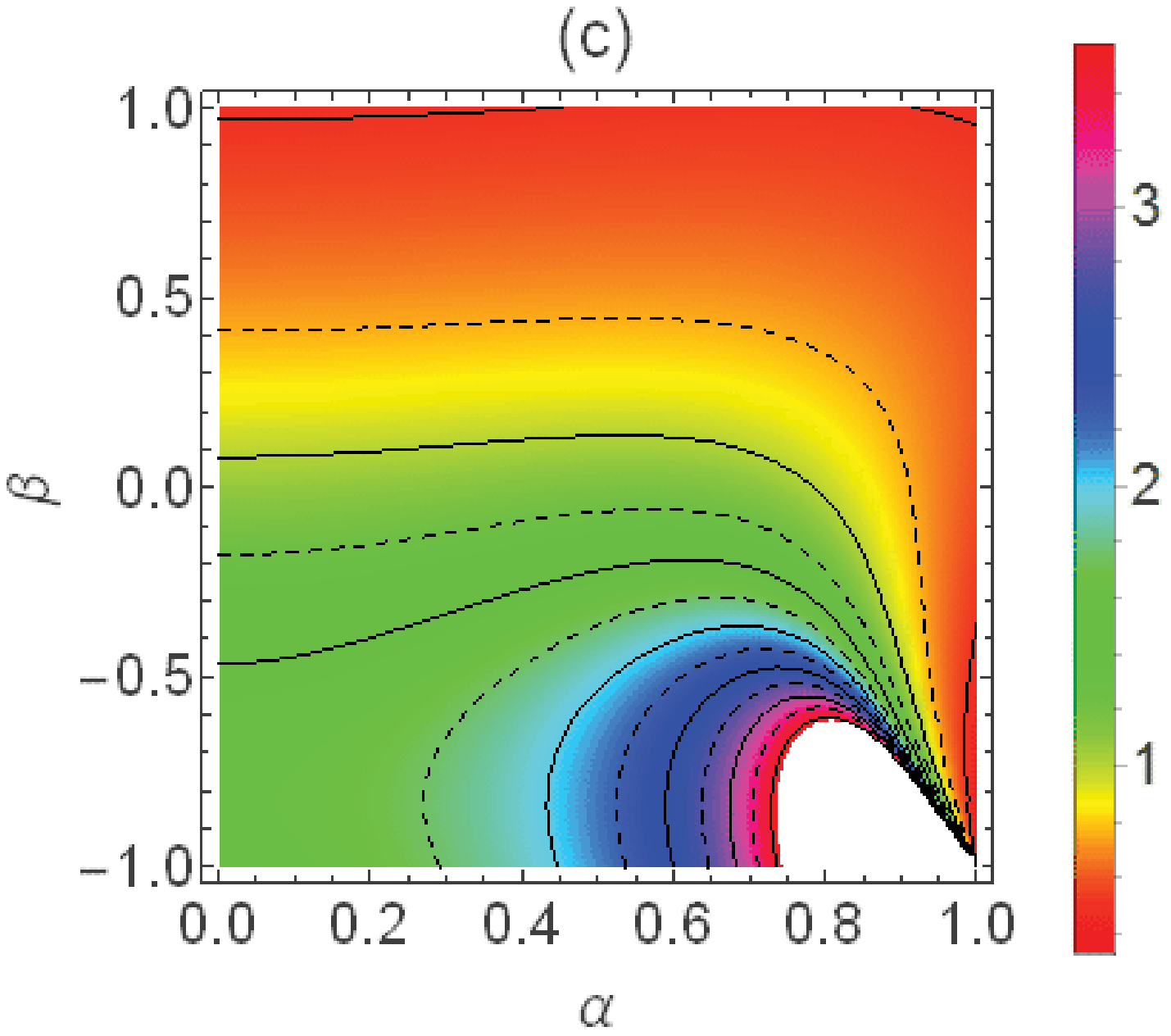}
\caption{(Color online) Same as in Fig.\ \protect\ref{fig2} but for the reduced bulk viscosity $\eta_b^*$. The contour lines  in panel (c) correspond to $\eta_b^*=0.5,0.75,\ldots,3.5$.}
\label{fig3}
\end{figure}

\begin{figure}
\includegraphics[width=8cm]{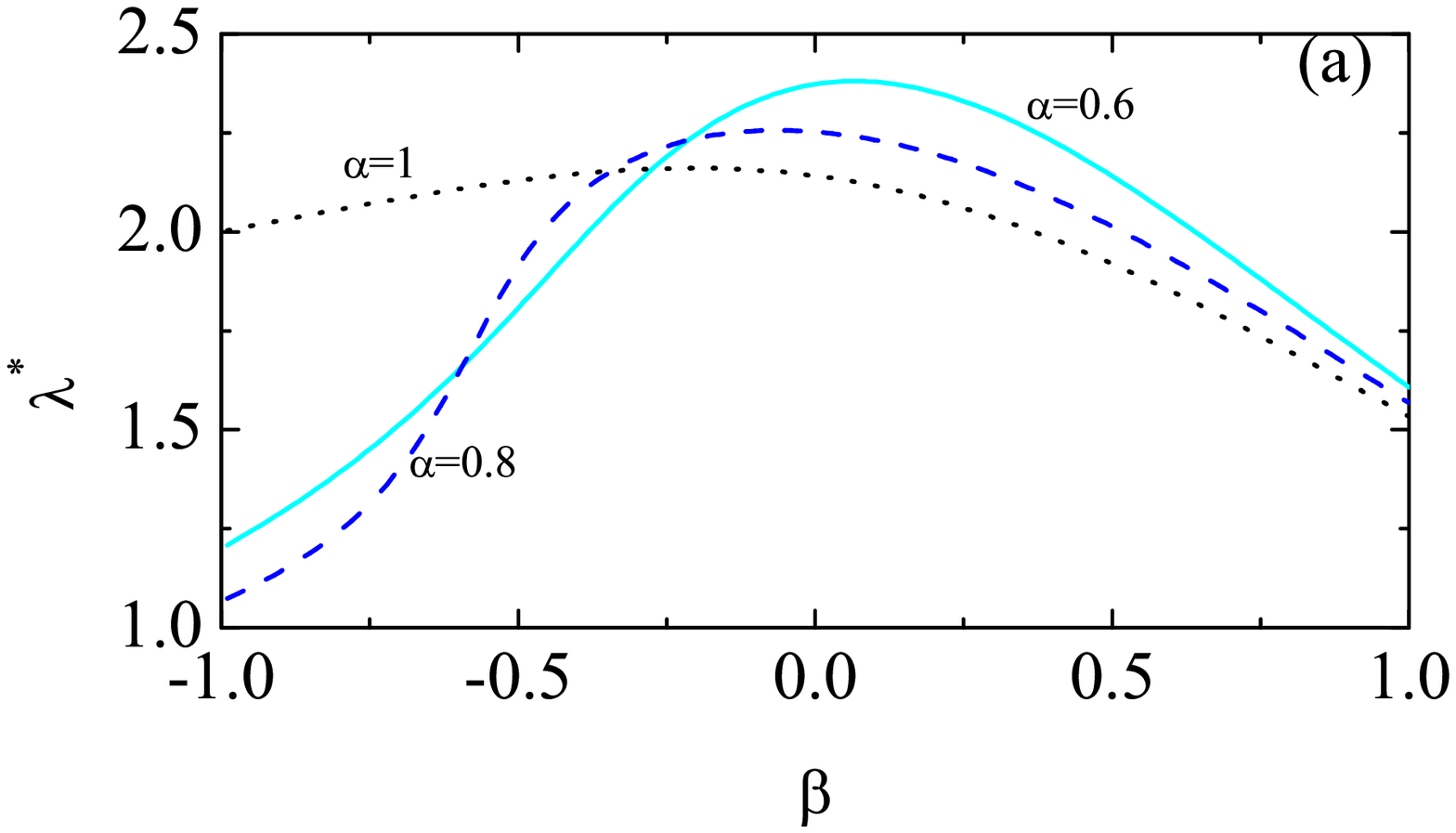}\\
\vskip0.2cm
\includegraphics[width=8cm]{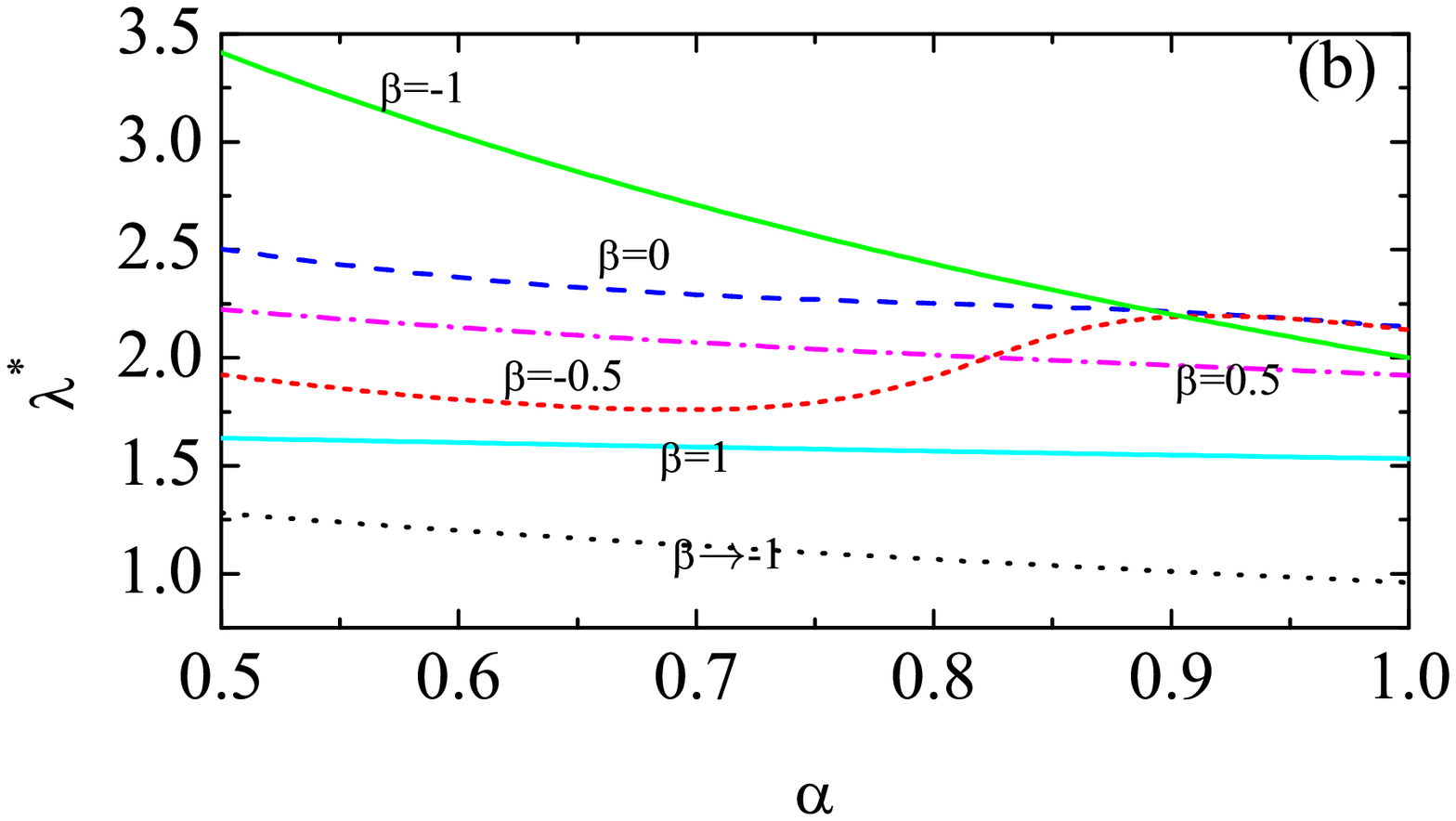}\\
\vskip0.2cm
\includegraphics[width=8cm]{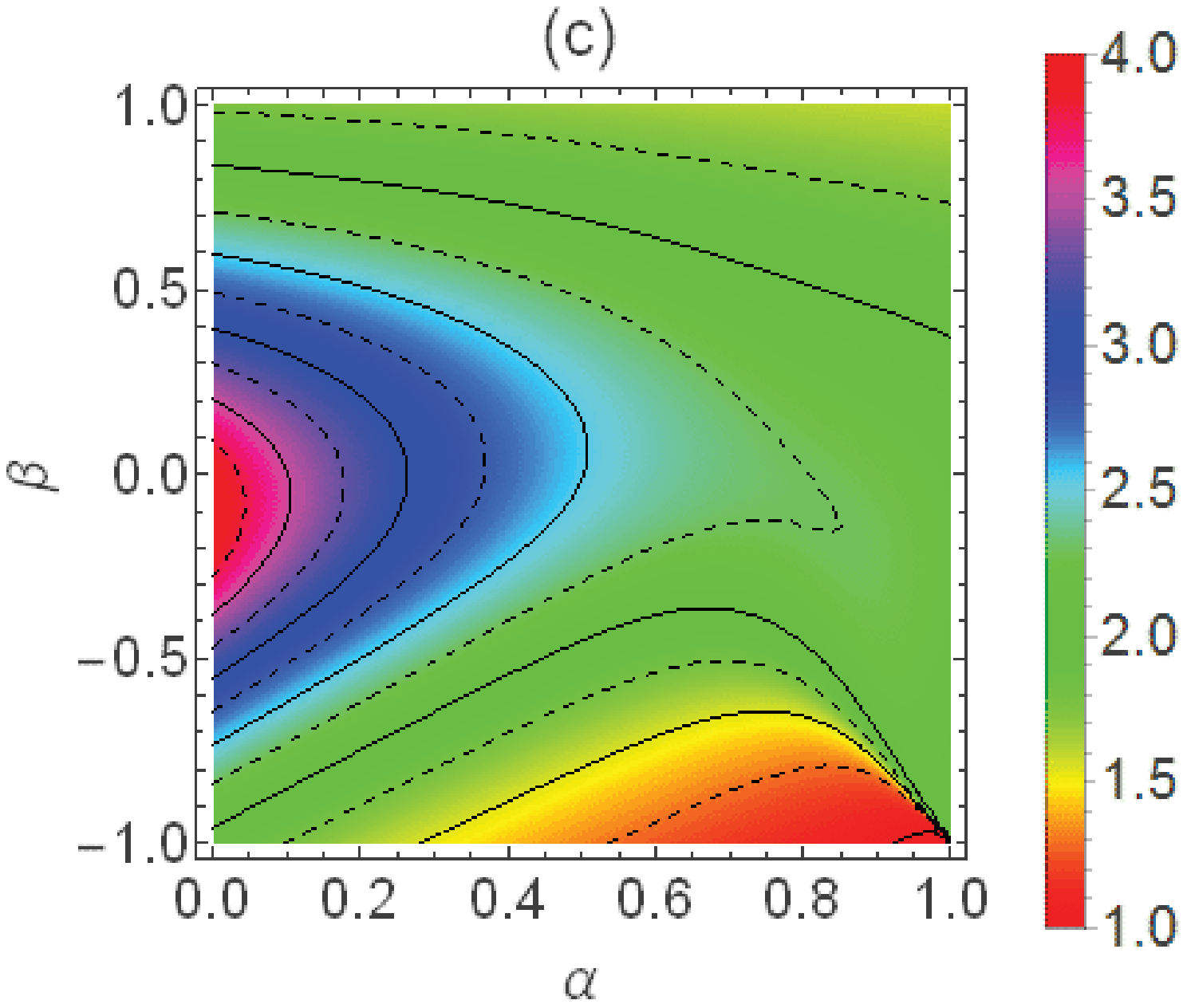}
\caption{(Color online) Same as in Fig.\ \protect\ref{fig2} but for the reduced thermal conductivity $\lambda^*$. The contour lines  in panel (c) correspond to $\lambda^*=1,1.25,\ldots,3.75$.}
\label{fig4}
\end{figure}

\begin{figure}
\includegraphics[width=8cm]{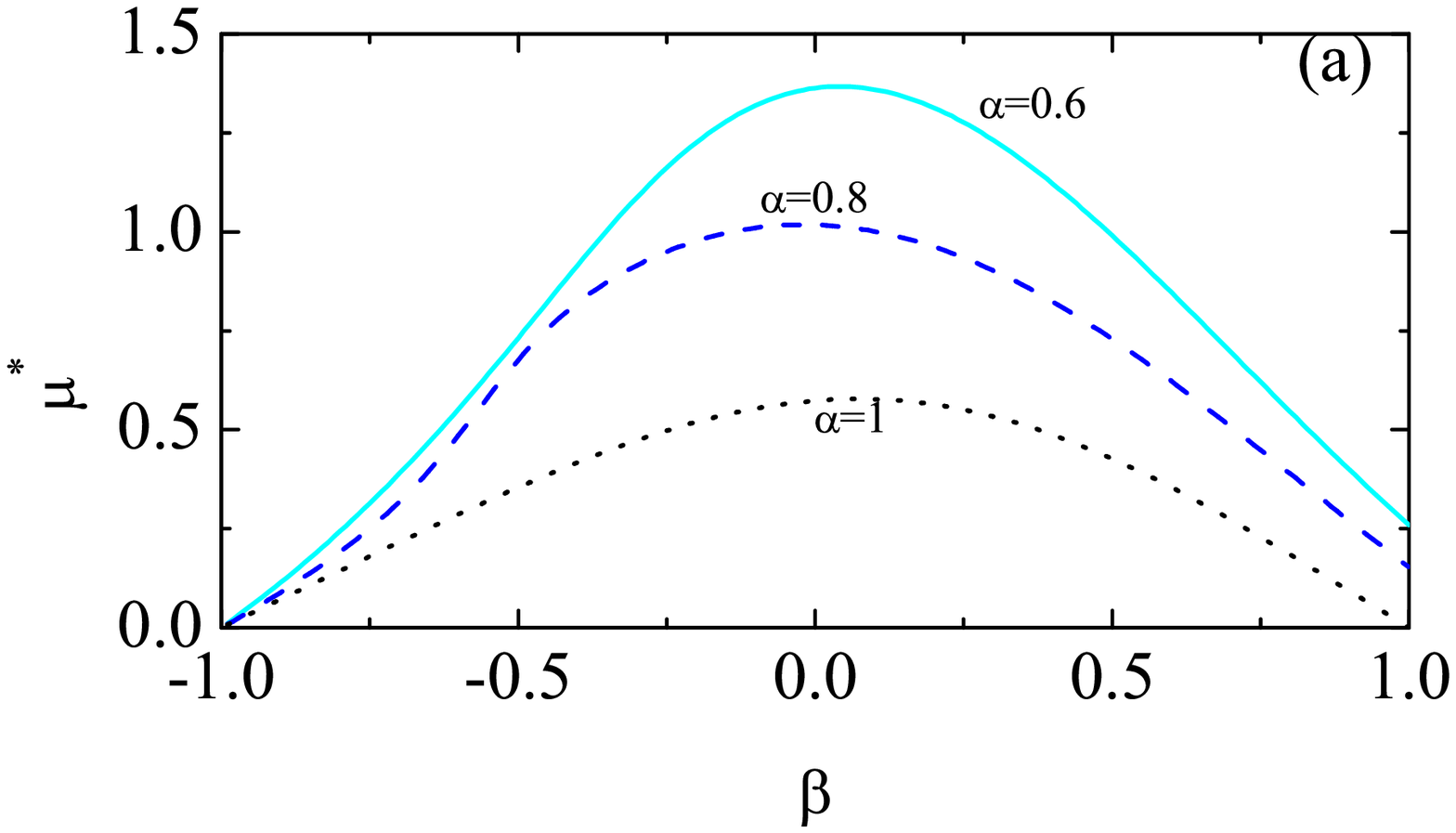}\\
\vskip0.2cm
\includegraphics[width=8cm]{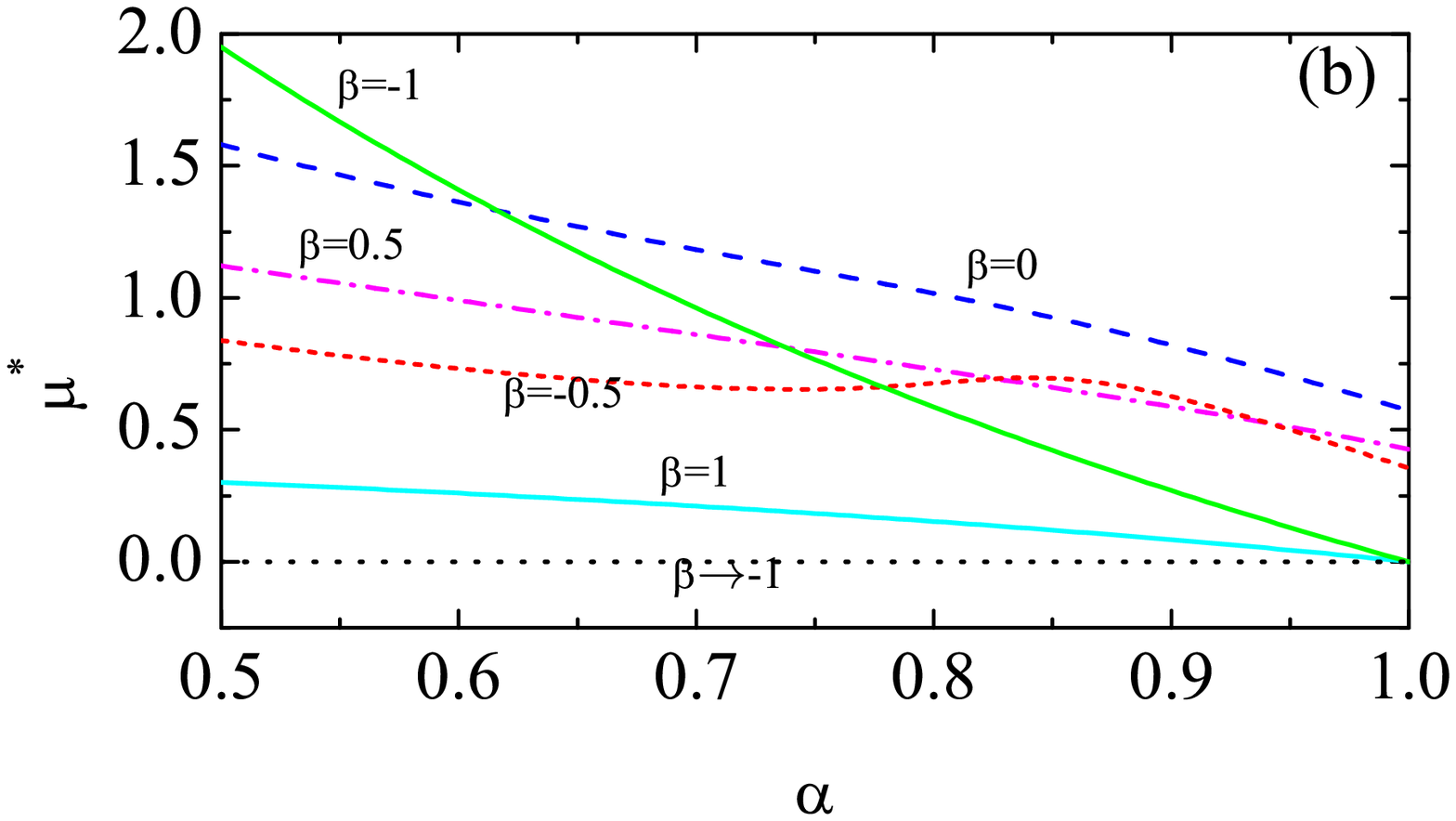}\\
\vskip0.2cm
\includegraphics[width=8cm]{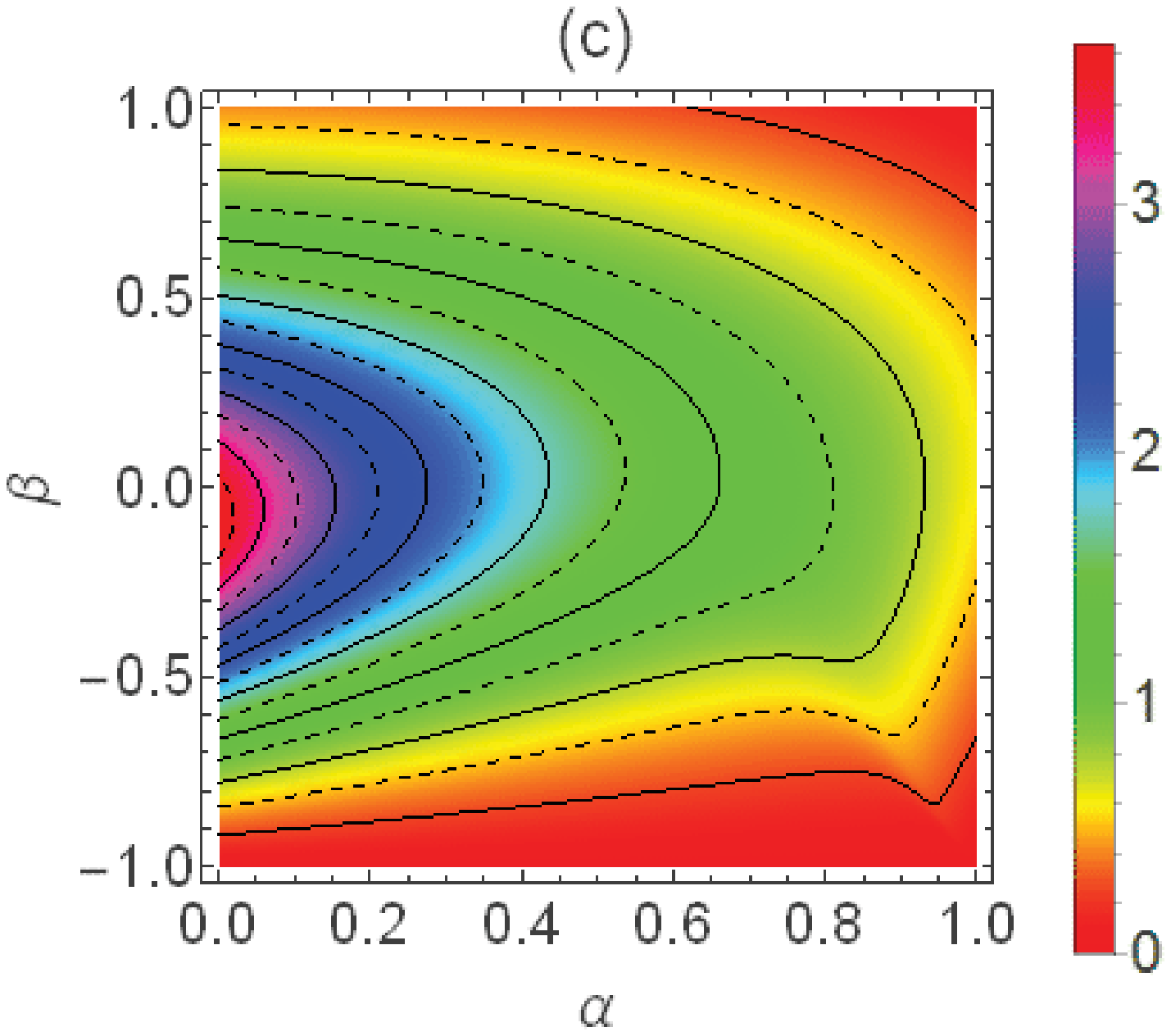}
\caption{(Color online) Same as in Fig.\ \protect\ref{fig2} but for the reduced Dufour-like coefficient $\mu^*$. The contour lines  in panel (c) correspond to $\mu^*=0.25,0.5,\ldots,3.5$.}
\label{fig5}
\end{figure}

\begin{figure}
\includegraphics[width=8cm]{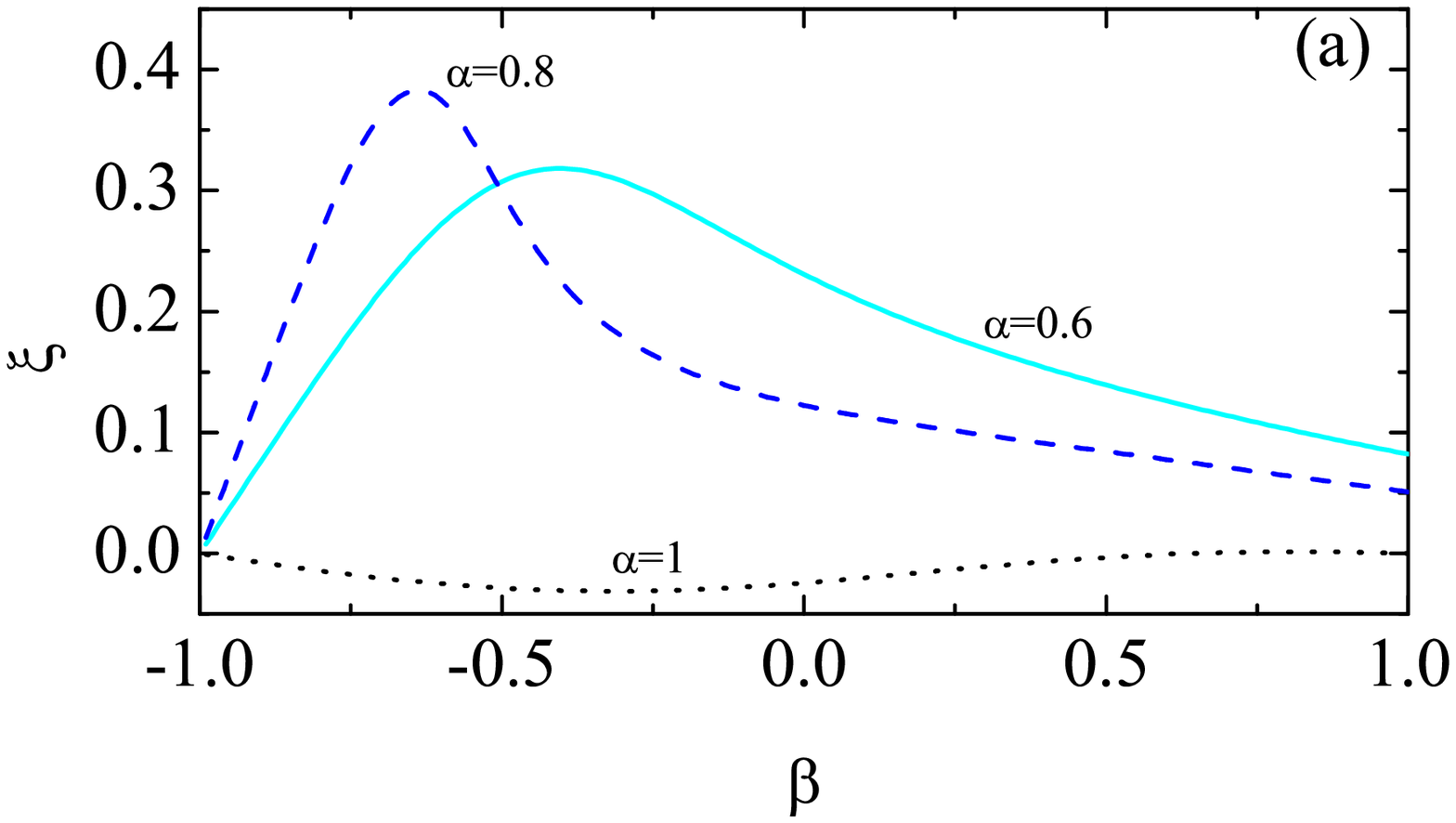}\\
\vskip0.2cm
\includegraphics[width=8cm]{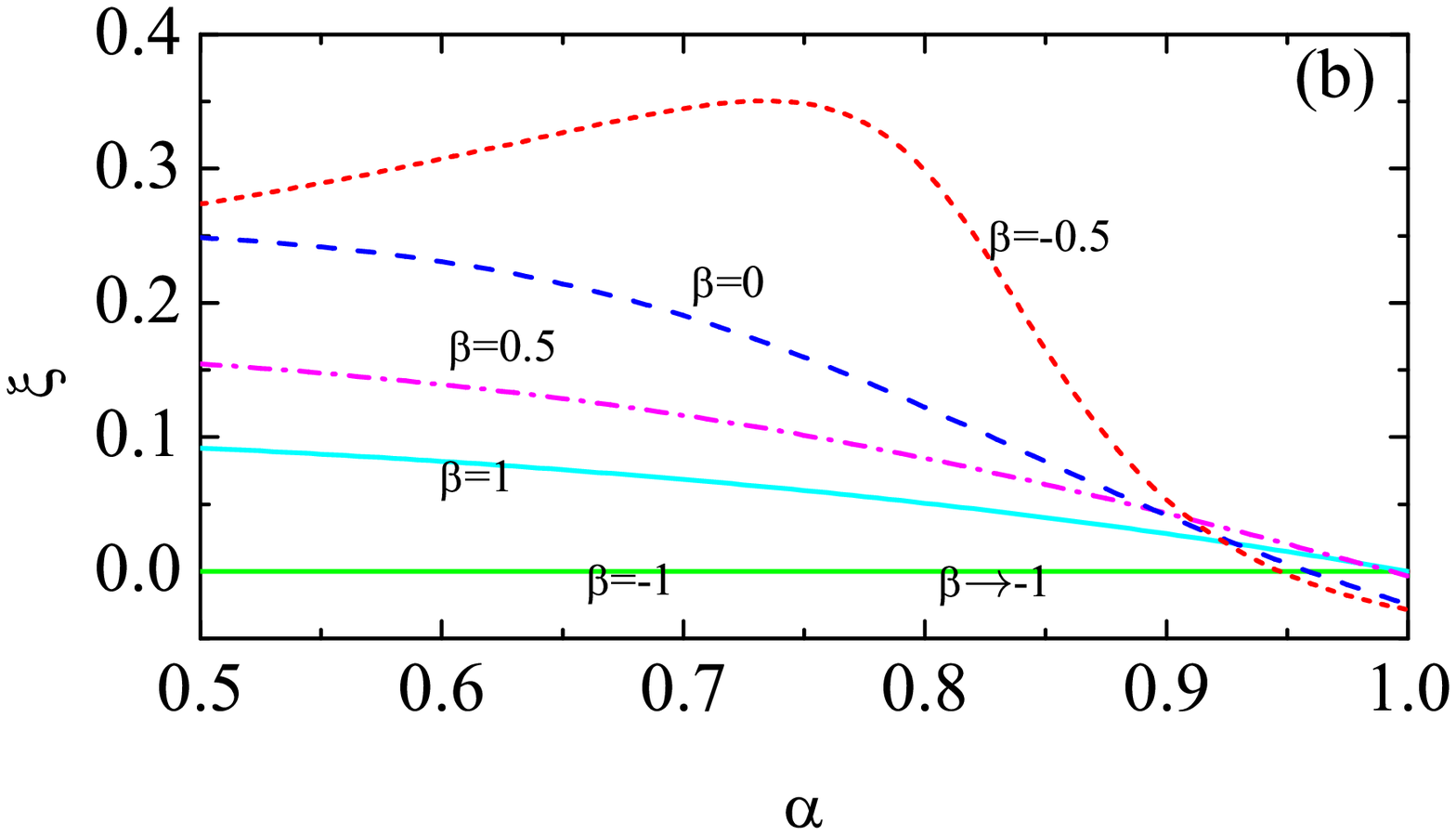}\\
\vskip0.2cm
\includegraphics[width=8cm]{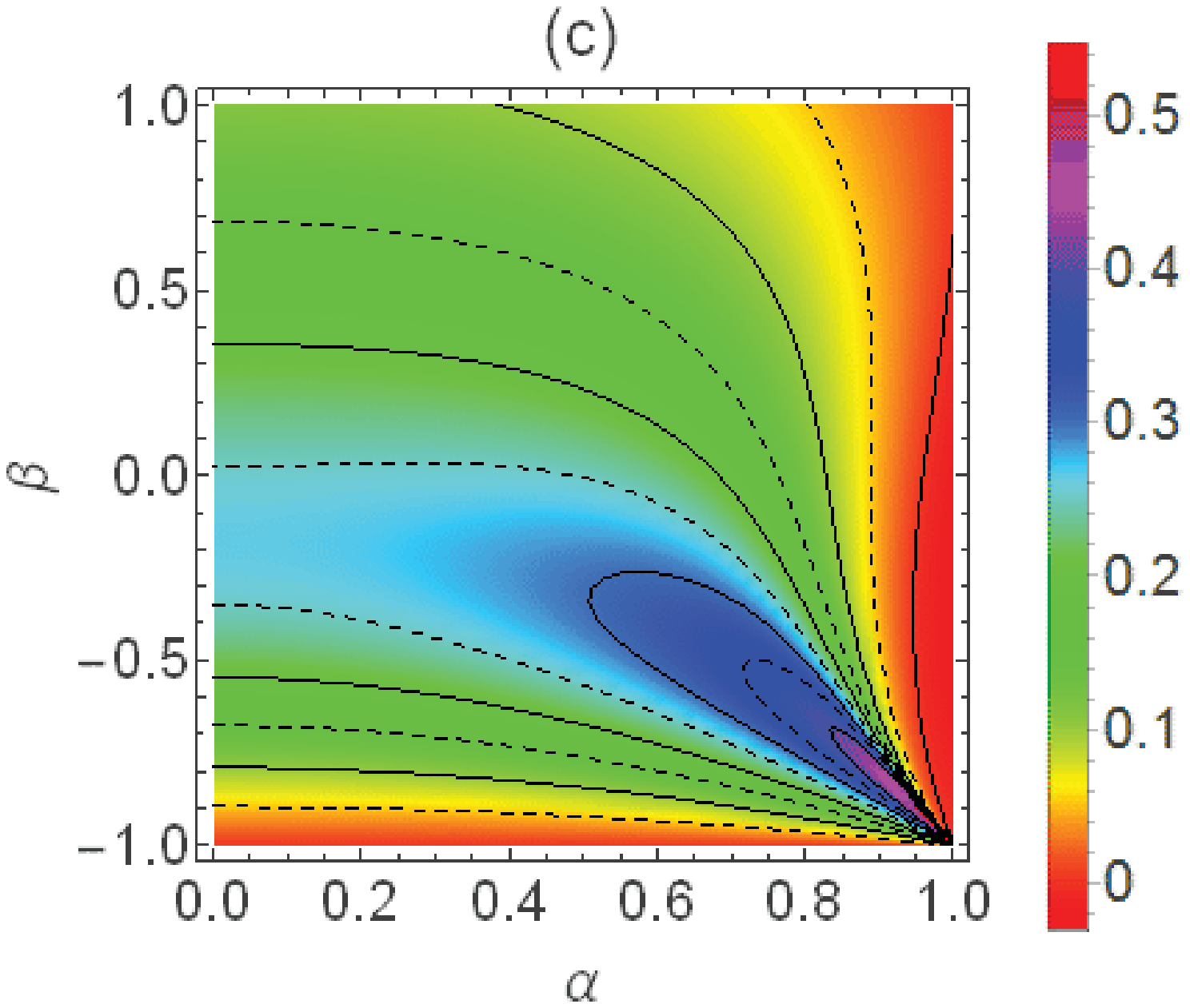}
\caption{(Color online) Same as in Fig.\ \protect\ref{fig2} but for the cooling rate transport coefficient $\xi$. The contour lines  in panel (c) correspond to $\xi=0,0.05,\ldots,0.4$.}
\label{fig6}
\end{figure}

\section{Discussion}
\label{sec6}
Let us now analyze the dependence of the five transport coefficients on both $\alpha$ and $\beta$. In order to define dimensionless quantities, we will take as a reference the transport coefficients (shear viscosity and thermal conductivity) of a gas made of elastic and smooth spheres at the same translational temperature $\tau_t T$ as that of the HCS of the granular gas, i.e.,
\beq
\eta_0=\frac{n\tau_t T}{\nu},\quad \lambda_0=\frac{15}{4}\frac{\eta_0}{m}.
\eeq
More specifically, the dimensionless shear and bulk viscosities are
\beq
\eta^*=\frac{\eta}{\eta_0},\quad \eta_b^*=\frac{\eta_b}{\eta_0},
\eeq
while the dimensionless thermal conductivity and Dufour-like coefficients are
\beq
\lambda^*=\frac{\lambda}{\lambda_0},\quad \mu^*=\frac{\mu}{\lambda_0}\frac{n}{ T}.
\eeq

\subsection{Limiting cases}
Before considering the case of general $\alpha$ and $\beta$, it is worth considering some limiting cases. We start with the case of a granular gas constituted by purely smooth inelastic hard spheres ($\beta=-1$ with arbitrary $\alpha$). In such a gas, the rotational degrees of freedom are irrelevant and should not contribute to the transport properties. A convenient way of isolating the relevant translational properties consists in formally setting $\thetah=0$ (i.e., $\tau_r=0$ and $\tau_t=2$) in the general expressions of Table \ref{table1},  apart from taking $\beta=-1$.
Alternatively, one can take $\thetah=0$ and $\kappa=0$, the latter defining spheres with a vanishing moment of inertia.
In both cases one obtains the results presented in Table \ref{table2}, which are consistent  with those previously derived for smooth inelastic hard spheres \cite{BDKS98,note_14_04_1}.

{\renewcommand{\arraystretch}{1.2}
\begin{table*}
   \caption{Special limiting cases.}\label{table2}
\begin{ruledtabular}
\begin{tabular}{cccc}
Quantity&Pure smooth&Quasismooth limit&Perfectly rough and elastic  \\
&($\b=-1$)&($\b\to -1$)&($\alpha=\beta=1$) \\ \hline
$\eta^*$&$\displaystyle{\frac{24}{(1+\a)(13-\a)}}$&$\displaystyle{\frac{24}{(1+\a)(19-7\a)}}$&$\displaystyle{\frac{6(1+\kappa)^2}{6+13\kappa}}$\\
$\eta_b^*$&$0$&$\displaystyle{\frac{8}{5(1-\a^2)}}$&$\displaystyle{\frac{(1+\kappa)^2}{10\kappa}}$\\
$\lambda^*$&$\displaystyle{\frac{64}{(1+\a)(9+7\a)}}$&$\displaystyle{\frac{48}{25(1+\a)}}$&$\displaystyle{\frac{12 (1+\kappa)^2 \left(37+151 \kappa +50 \kappa ^2\right)}{25 \left(12+75 \kappa +101 \kappa ^2+102 \kappa^3\right)}}$\\
$\mu^*$&$\displaystyle{\frac{1280(1-\a)}{(1+\a)(9+7\a)(19-3\a)}}$&$0$&$0$\\
$\xi$&$0$&$0$&$0$\\
 \end{tabular}
 \end{ruledtabular}
 \end{table*}
}

Next, we consider the quasismooth limit $\beta\to -1$. The results differ from those obtained before for pure smooth spheres because, {as shown in Sec.\ \ref{sec3}},  the HCS rotational-translational temperature ratio diverges as $\thetah\sim (1+\beta)^{-2}$, and thus the rotational contributions cannot be neglected. The final expressions, which are independent of the reduced moment of inertia $\kappa$, are also included in Table \ref{table2}. In contrast, it can be checked that the expressions in the limit of small moment of inertia ($\kappa\to 0$) depend on $\beta$.

Finally, let us analyze the physically important case of perfectly elastic and rough spheres ($\alpha=\beta=1$). This defines a conservative system (i.e., the total kinetic energy is conserved by collisions) that has been used for a long time to model polyatomic gases \cite{P22,MSD66,CC70,K10a}. The results obtained in this limit are displayed in the last column of Table \ref{table2} and  fully agree with those first derived by Pidduck \cite{P22}.

\subsection{General $\alpha$ and $\beta$}

Now we go back to a granular gas with general values of $\alpha$, $\beta$, and $\kappa$, in which case the expressions for the five transport coefficients are given in Table \ref{table1}. For the sake of concreteness, let us restrict ourselves to spheres with a uniform mass distribution, so that $\kappa=\frac{2}{5}$.

Figures \ref{fig2}--\ref{fig6} show the dependence of the reduced transport coefficients $\eta^*$, $\eta_b^*$, $\lambda^*$, $\mu^*$, and $\xi$, respectively, on both $\alpha$ and $\beta$. {As in Fig.\ \ref{nfig3},} in the top panels the quantities are plotted versus $\beta$ for three representative values of the coefficient of normal restitution ($\alpha=0.6$, $0.8$, and $1$). The middle panels present the dependence on $\alpha$ for a few representative values of the coefficient of tangential restitution, namely $\beta=-0.5$, $0$, $0.5$, and $1$. Additionally, the quasismooth limit ($\beta\to -1$) and the case of purely smooth spheres ($\beta=-1$) are also considered. Finally, the bottom panels represent density plots in the $\alpha$-$\beta$ plane.

Let us start analyzing the three transport coefficients that are also present in the case of purely smooth particles, i.e.,  the (reduced) shear viscosity, thermal conductivity, and Dufour-like coefficient (Figs.\ \ref{fig2}, \ref{fig4}, and \ref{fig5}, respectively).  We observe that, at fixed $\alpha$, those coefficients present a non-monotonic $\beta$ dependence with  maxima around $\beta\approx 0$. On the other hand,  the dependence on $\alpha$ is rather sensitive to the value of $\beta$, showing an intricate interplay between both coefficients of restitution. Typically, the transport coefficients increase with increasing inelasticity, although some exceptions are found (see, for instance, $\eta^*$ at $\beta=1$ and $\lambda^*$ and $\mu^*$ at $\beta=-0.5$). Furthermore, an interesting observation from Figs.\ \ref{fig2}, \ref{fig4}, and \ref{fig5} is that the impact of the coefficient of normal restitution on the transport coefficients $\eta^*$, $\lambda^*$, and $\mu^*$ is much milder in the case of rough spheres than for purely smooth spheres.
{Comparison between Fig.\ \ref{nfig3}, on the one hand, and Figs.\ \ref{fig2}, \ref{fig4}, and \ref{fig5}, on the other hand, shows that the general dependencies of the transport coefficients $\eta^*$, $\lambda^*$, and $\mu^*$ on both $\alpha$ and $\beta$ (in particular, the maxima around $\beta\sim 0$) are highly correlated to that of the cooling rate $\zeta^*$. This explains that  fair \emph{qualitative} estimates can be obtained by using the expressions for smooth spheres \cite{BDKS98,GD99} with the cooling rate replaced by the one for rough spheres \cite{MDHEH13}. }

Next, we consider the two transport coefficients that vanish for purely smooth particles. As observed from Fig.\ \ref{fig3}, the bulk viscosity $\eta_b^*$ exhibits a highly nontrivial behavior. It reaches especially high values in the quasielastic and quasismooth region, diverging in the limit $\alpha\to 1$, $\beta\to -1$. Outside that region, the bulk viscosity can be larger than the shear viscosity. For instance, $\eta_b^*/\eta^*\simeq 1.26$ at $\alpha=0.7$ and $\beta=-0.2$. As for the cooling rate transport coefficient $\xi$ (see Fig.\ \ref{fig6}), it also presents a complex behavior.  A remarkable feature is that it becomes negative in a certain region of the $\alpha$-$\beta$ plane near $\alpha=1$. {Of course, this does not mean that the cooling rate itself is negative or signals any breakdown of the Chapman--Enskog method, the Sonine approximation, or the friction model used. According to Eqs.\ \eqref{23c} and \eqref{zeta1xi}, a negative value of the transport coefficient $\xi$ simply implies that   the cooling rate $\zeta$ is larger (smaller) than its HCS value $\zeta^{(0)}$ if $\bnabla\cdot \mathbf{u}$ is positive (negative).}

\section{Concluding remarks}
\label{sec7}
In this work we have developed a hydrodynamic theory for a dilute granular gas modeled as a system of identical inelastic and rough hard spheres. Energy dissipation in collisions is characterized by two constant coefficients of restitution: the normal ($0<\alpha\leq 1$) and the tangential ($-1\leq\beta\leq 1$) coefficients. In this model both the translational velocity ($\mathbf{v}$) of the center of mass and the angular velocity ($\bw$) of the particles are mutually influenced by collisions.

The methodology has been based on the Boltzmann kinetic equation for the one-particle velocity distribution function $f(\bx,\bc,\bw,t)$. The kinetic equation has been solved by means of the Chapman--Enskog method \cite{CC70,FK72} for states with small spatial gradients of the hydrodynamic fields: the number density $n(\bx,t)$, the flow velocity $\mathbf{u}(\bx,t)$, and the (total) granular temperature $T(\bx,t)$. The solution provides the constitutive equations for the pressure tensor $P_{ij}(\bx,t)$, the heat flux $\mathbf{q}(\bx,t)$, and the cooling rate $\zeta(\bx,t)$. The associated five transport coefficients (shear viscosity $\eta$, bulk viscosity $\eta_b$, thermal conductivity $\lambda$, Dufour-like coefficient $\mu$, and cooling rate transport coefficient $\xi$) are exactly expressed in terms of integrals involving the solutions of a set of linear integral equations. In particular, the existence of $\eta_b$ and $\xi$ imply that, in the case of a compressible flow (i.e., $\bnabla\cdot\mathbf{u}\neq 0$), the rotational-translational temperature ratio and the cooling rate differ from their forms in the reference HCS.

As happens in the conventional case of elastic particles \cite{CC70}, explicit expressions for the transport coefficients can be obtained by expanding the zeroth- and first-order distributions in Sonine polynomials and truncating the expansions at the simplest level (the so-called first Sonine approximation). This has allowed us to determine the five transport coefficients as nonlinear functions of $\alpha$, $\beta$, and the reduced moment of inertia $\kappa$. For easy reference, the final results are displayed in Table \ref{table1}.

Our results extend to arbitrary values of $\alpha$ and $\beta$ previous works for inelastic and purely smooth spheres ($\alpha<1$, $\beta=-1$) \cite{BDKS98} and elastic and perfectly rough spheres ($\alpha=\beta=1$) \cite{P22,CC70,K10a}, as shown in Table \ref{table2}. Due to the coupling between translational and rotational degrees of freedom in the HCS, the quasismooth limit $\beta\to -1$ yields results differing from those for purely smooth spheres.

As Figs.\ \ref{fig2}--\ref{fig6} clearly show, the dependence of the transport coefficients on both $\alpha$ and $\beta$ is rather intricate.
On the other hand, it must be noted that, since the deviations of the reference HCS from the two-temperature Maxwellian \eqref{Maxw} are important in the region $-1\leq\beta\lesssim -0.5$ only \cite{VSK14},  the expressions derived here are expected to be especially reliable in the region $-0.5\lesssim \beta\leq 1$, which is likely the one of practical interest from an experimental point of view \cite{L99}.
{A more rigorous hydrodynamic theory in the region of small roughness ($\beta\gtrsim -1$) would likely require, apart from accounting for strong non-Maxwellian features of the HCS distribution, the inclusion of the mean spin $\boldsymbol{\Omega}$ as an additional hydrodynamic variable.}

The present work opens new challenges to explore. First, we plan to carry out a linear stability analysis \cite{G05} of the NSF hydrodynamic equations  to determine the critical length $L_c$ beyond which the HCS becomes unstable and assess the impact of roughness on $L_c$ \cite{MDHEH13}. Given that most of the experimental setups consider granular systems confined in two dimensions, we  intend to determine the NSF transport coefficients for systems of inelastic rough hard disks by using a methodology similar to the one followed here. Moreover, the structure of the collisional frequencies derived in  the Appendix can be exploited to obtain the NSF transport coefficients of driven granular gases, in analogy to the case of smooth spheres \cite{GM02,GMT13,GCV13}. Finally, we will test the transport coefficients obtained from the first Sonine approximation against DSMC numerical solutions  of the Boltzmann equation  by methods similar to those employed for smooth spheres \cite{GM02,MG03b,GM03,BR04,MSG05,BRMG05,MSG07}.

\begin{acknowledgments}
The research of A.S. and V.G.  was supported by the Spanish Government through Grant No.\ FIS2013-42840-P and  by the Junta de Extremadura (Spain) through Grant No.\ GR10158, both partially financed by FEDER funds.
The work of G.M.K. has been supported by the Conselho Nacional de Desenvolvimento Cient\'ifico e Tecnol\'ogico (Brazil).
\end{acknowledgments}

\appendix*
\section{Explicit expressions in the Sonine approximation}

This Appendix provides the steps needed to determine the transport coefficients within the Sonine approximations \eqref{Aapprox}--\eqref{Eapprox}. The task requires the evaluation of the collision integrals appearing in the collision frequencies \eqref{nulambdat}, \eqref{nulambdar}, \eqref{numut}, \eqref{numur}, and \eqref{nueta}.
The algebra involved in those collision integrals is rather tedious, so here we only provide the final results.

First, the collision frequency associated with the shear viscosity turns out to be
\beq
\nu_\eta^*\equiv \frac{\nu_\eta}{\nu}=(\at+\bt)(2-\at-\bt)+\frac{\bt^2\thetah}{6\kappa},
\label{etaetab}
\eeq
where $\at$ and $\bt$ are defined by Eq.\ \eqref{7} and the HCS temperature ratio $\thetah$ is given by Eqs.\ \eqref{1.8} and \eqref{1.9}. The shear viscosity coefficient is directly obtained from Eq.\ \eqref{eta}.

Now we consider the collision integrals  \eqref{xitr}. Insertion of Eq.\ \eqref{Eapprox} gives
\beq
\xi_t=\gamma_E\Xi_t,\quad  \xi_r=\gamma_E\Xi_r,
\label{Xitr}
\eeq
with
\beq
\Xi_t=\frac{5}{8}\tau_r\Big[1-\a^2+(1-\b^2)\frac{\kappa}{1+\kappa}-\frac{\kappa}{3}({\thetah-5})\left(\frac{1+\b}{1+\kappa}\right)^2\Big],
\eeq
\beq
\Xi_r=\frac{5}{8}\tau_t\frac{1+\beta}{1+\kappa}\left[\frac{\thetah-2}{3}(1-\beta)
+\frac{\kappa}{3}({\thetah-5})\frac{1+\b}{1+\kappa}\right].
\label{A4}
\eeq
Combination of Eqs.\ \eqref{etab}, \eqref{X1},  and \eqref{Xitr} yields
\beq
\gamma_E=\frac{2}{3}\frac{1}{\Xi_t-\Xi_r-\zeta^*}.
\eeq
This closes the evaluation of the bulk viscosity $\eta_b$. Moreover, the cooling rate coefficient $\xi$ defined by Eq.\ \eqref{zeta1xi} is, according to Eqs.\ \eqref{xi} and \eqref{Xitr}--\eqref{A4},
\beq
\xi=\frac{5}{16}\tau_t\tau_r\gamma_E\left[1-\a^2+\left(1+\frac{1}{3}\frac{\thetah-5}{1+\kappa}\right)(1-\b^2)\right].
\eeq

Next, we turn our attention to the heat flux coefficients. The collision frequencies $\nu_{\lambda_t}$, $\nu_{\lambda_r}$, $\nu_{\mu_t}$, and $\nu_{\mu_r}$ turn out to be given by
\beq
\nu_{\lambda_t}^*\equiv\frac{\nu_{\lambda_t}}{\nu}=Y_t+Z_t \frac{\gamma_{A_r}}{\gamma_{A_t}},
\label{A7}
\eeq
\beq
\nu_{\lambda_r}^*\equiv\frac{\nu_{\lambda_r}}{\nu}=Y_r\frac{\gamma_{A_t}}{\gamma_{A_r}}+Z_r,
\label{A8}
\eeq
\beq
\nu_{\mu_t}^*\equiv\frac{\nu_{\mu_t}}{\nu}=Y_t+Z_t \frac{\gamma_{B_r}}{\gamma_{B_t}},
\label{A9}
\eeq
\beq
\nu_{\mu_r}^*\equiv\frac{\nu_{\mu_r}}{\nu}=Y_r\frac{\gamma_{B_t}}{\gamma_{B_r}}+Z_r,
\label{A10}
\eeq
with
\beq
Y_t=\frac{41}{12}\left(\at+\bt\right)-\frac{33}{12}\left(\at^2+\bt^2\right)-\frac{4}{3}\at\bt-\frac{7\thetah}{12}
\frac{\bt^2}{\kappa},
\eeq
\beq
Z_t=-\frac{5\thetah}{6}
\frac{\bt^2}{\kappa},
\eeq
\beq
Y_r=\frac{25}{36}\left(\frac{\bt}{\kappa}-3\frac{\bt^2}{\thetah\kappa}-\frac{\bt^2}{\kappa^2}\right),
\eeq
\beq
Z_r=\frac{5}{6}\left(\at+\bt\right)+\frac{5}{18}\frac{\bt}{\kappa}\left(7-3\frac{\bt}{\kappa}-6{\bt}-
 4{\at}\right).
 \eeq
{}From Eqs.\ \eqref{lambdat}, \eqref{lambdar}, \eqref{X2}, \eqref{A7}, and \eqref{A8} one obtains a set of two algebraic linear equations for $\gamma_{A_t}$ and $\gamma_{A_r}$ whose solution is
\beq
\gamma_{A_t}=\frac{Z_r-Z_t-2\zeta^*}{\left(Y_t-2\zeta^*\right)\left(Z_r-2\zeta^*\right)-Y_r Z_t},
\label{A15}
\eeq
\beq
\gamma_{A_r}=\frac{Y_t-Y_r-2\zeta^*}{\left(Y_t-2\zeta^*\right)\left(Z_r-2\zeta^*\right)-Y_r Z_t}.
\label{A16}
\eeq
Upon deriving these equations we have made use of the approximation $a_{20}^{(0)}=a_{11}^{(0)}=0$, in consistency with Eq.\ \eqref{Maxw}.
Equations \eqref{A15} and \eqref{A16}, together with Eq.\ \eqref{X2}, close the determination of the thermal conductivity coefficients $\lambda_t$ and $\lambda_r$.

Analogously, once the coefficients $\gamma_{A_t}$ and $\gamma_{A_r}$ are known, Eqs.\ \eqref{mut}, \eqref{mur}, \eqref{X3}, \eqref{A9}, and \eqref{A10} yield a set of two linear equations with the solution
\beq
\gamma_{B_t}=\zeta^*\frac{\gamma_{A_t}\left(Z_r-\frac{3}{2}\zeta^*\right)-\gamma_{A_r}Z_t}
{\left(Y_t-\frac{3}{2}\zeta^*\right)\left(Z_r-\frac{3}{2}\zeta^*\right)-Y_r Z_t},
\eeq
\beq
\gamma_{B_r}=\zeta^*\frac{\gamma_{A_r}\left(Y_t-\frac{3}{2}\zeta^*\right)-\gamma_{A_t}Y_r}
{\left(Y_t-\frac{3}{2}\zeta^*\right)\left(Z_r-\frac{3}{2}\zeta^*\right)-Y_r Z_t}.
\eeq
This closes the evaluation of the Dufour-like coefficients $\mu_t$ and $\mu_r$.

\bibliographystyle{apsrev}

\bibliography{D:/Dropbox/Public/bib_files/Granular}

\begin{thebibliography}{71}
\bibitem[{not({\natexlab{c}})}]{note_14_04_1}
\bibinfo{note}{Note that there is an extra factor $2$ in the expressions of
  $\lambda^*$ and $\mu^*$, as compared with those in Ref.\
  \protect\cite{BDKS98}. This is due to the fact that the translational
  temperature is actually $2T$ in the smooth-sphere limit.}
\end{thebibliography}

\end{document}